\documentclass[11pt]{article}

\usepackage[T1]{fontenc}
\usepackage[utf8]{inputenc}
\usepackage[margin=1in]{geometry}
\usepackage{amsmath}
\usepackage{amssymb}
\usepackage{booktabs}
\usepackage{fancyvrb}
\usepackage[numbers,sort&compress]{natbib}
\usepackage{microtype}
\usepackage[htt]{hyphenat}
\usepackage[hidelinks]{hyperref}

\newtheorem{definition}{Definition}

\newcommand{\vacname}{VAC}

\title{Four Ways to Forge a Bundle My Own Verifier Calls Clean\\
\large Refusal-Site Mutation Testing of an Evidence-Bundle Verifier}

\author{%
  Erik Hill\\
  Independent Researcher\\
  \texttt{erik@erikhill.dev}\\
  ORCID \href{https://orcid.org/0009-0002-5912-967X}{0009-0002-5912-967X}\\
  \url{https://erikhill.dev}
}

\date{18 August 2026}

\begin{document}
\maketitle

\begin{abstract}
I built a protocol whose premise is that a stranger can re-run my claims offline and get
the same answer. An outside engineer audited it and broke it: a bundle whose
headline numbers were false verified clean, the cheapest forgery four bytes. I merged his fix, then pointed my own
instruments at the fixed verifier and found the same defect four more times, in places his
audit did not reach. The cheapest is one capital letter.

The unifying defect is not cryptographic or exotic: a check that reports success along a
path where it never examined anything. \emph{Vacuous pass} is a working label, not a
discovery; Section~\ref{sec:related} names the literatures already occupying it.

So I stopped collecting anecdotes and measured. At \texttt{f59fb62}, under the extraction
rule of Section~\ref{sec:method}, the verifier exposes 112 refusal sites; 75 could be deleted with
the whole suite and every tamper fixture still green, a score of 0.330. Three of the four hand-found forgeries fall in surviving classes; the fourth is an obligation
with no refusal site. Scored alone, the sixteen-fixture corpus built to prove
the verifier can refuse catches 10. Testing
the refusals themselves took it to 0.941, then to 1.000 at \texttt{92e4548} over a grown
population of 146 sites; those denominators differ and the series between them is non-monotone, so
Section~\ref{sec:liveness-tests} carries all eleven, not just the five rows of
Table~\ref{tab:ladder}. Fixing the four
found defects instead moved 37/112 to 39/119, leaving the pre-existing sites at 37.

Seven times during this study my own measuring tools reported success while measuring
nothing; four were built to detect this class, and one returned a perfect 1.000.

Every number here is self-measured on a system I wrote, over a registry that is a closed loop
of my own repositories; the one external data point is the audit of Section~\ref{sec:audit}.
That is stated here rather than buried: it is the paper's credibility, not a caveat.
\end{abstract}


\section{Introduction}
\label{sec:intro}

A verifier is a grader. Its output is a verdict that other people act on. When it accepts,
the thing it accepted is copied into a public registry and rendered on a published page.
So the interesting question about a verifier is not whether it accepts good input. It is
whether its refusals can fire at all.

This paper is a case study of one verifier: mine. It began when an outside engineer
submitted a bundle whose headline numbers were false, with every hash pinned honestly, and
my verifier printed \texttt{structural verification: PASS}. The recomputation had run over
an empty artifact, found no mismatch, and passed. The forgery cost four bytes.

I merged the fix. Then I asked a narrower question than ``are there other bugs'': how much
of this verifier's acceptance logic has ever been observed rejecting anything? To answer it
I disabled the verifier's refusal statements one at a time and asked whether the test suite,
a liveness control, or a committed corpus of tampered bundles noticed. Two thirds of them
could be deleted with everything still green.

\subsection*{Research questions}

\begin{description}
\item[RQ1] What fraction of the verifier's refusal sites can be deleted with the full unit
suite, the liveness control, and the committed tamper corpus all still green?
\item[RQ2] Does regression testing the specific defects an external audit found improve
that fraction?
\item[RQ3] Does systematic per-refusal liveness testing improve it?
\item[RQ4] Does the operator's classification of surviving refusals anticipate forgeries
found independently by hand?
\end{description}

Every one of these has a measured answer, and one of them is a null result I would not have
predicted.

\subsection*{Contributions}

\begin{enumerate}
\item A reproducible case study of five fail-open and vacuous-pass classes in an offline
evidence-bundle verifier, each with a working forgery: the four in
Section~\ref{sec:forgeries}, found against a verifier that had already been hardened once,
plus the \texttt{null}-artifact forgery in Section~\ref{sec:audit} reported by the external
audit.
\item A refusal-site liveness instrument: a deletion operator applied to an enumerated
population of the verifier's own refusal statements, scored by a deterministic two-detector
kill predicate, with an explicit exclusion set and a validity gate. Its operator, its purpose
and its precondition each have a published parent. Deleting the statement that signals
rejection is Kumar et al.'s; grading a checker by mutation is muSE's, MASC's, Chen and
Furia's and, concurrently, Delcourt et al.'s; the green-baseline precondition is standard
mutation-tool behaviour. What is offered is the pairing set out in
Sections~\ref{sec:related-neighbours} and~\ref{sec:related-unclaimed}: the site selection and
the subject, applied to the checker's own source rather than to the artifact it reads. The
kill predicate is what makes the measurement decomposable, not what makes it new.
\item An empirical before-and-after on one real verifier, comparing discovered-bug
regression testing against systematic refusal-site liveness testing. The first moved
coverage from $37/112$ to $39/119$ and left the caught set on the pre-existing sites at
exactly 37. The second moved it to $112/119$. The two arms are not effort-matched, the
second is measured on top of the first rather than as an alternative to it, and
Section~\ref{sec:liveness-tests} states why the second arm's gain is partly entailed by
how the intervention is defined. The null result in the first arm was independently
anticipated, on a system I did not write, by Bilal and Mughal
(Section~\ref{sec:related-neighbours}).
\item A separately scored detector result. The committed corpus of tampered bundles, written
for the express purpose of proving the verifier can refuse, catches 10 of 112 refusal sites
on its own, against a structural ceiling of $16/112$ that no amount of care in writing those
sixteen fixtures could beat (Section~\ref{sec:rq1}). MASC reports the same shape of result
from the other side of the pipeline, and Section~\ref{sec:related-unclaimed} sets out how
the two partitions differ.
\item A finding-shaped hypothesis, untested outside this codebase, that an evidence artifact
emitted as a report is systematically underbound relative to one consumed as a check's input
(Section~\ref{sec:closure}). The narrower form, keyed to whether the artifact is human-readable,
is refuted there by this paper's own data at $n=4$.
\item Seven documented cases in which the measuring instruments reported success while
measuring nothing (Section~\ref{sec:instruments}), including one that returned a perfect
score. None of the three closest neighbouring papers reports an instrument of its own
returning a passing verdict along a path where it performed no comparison, although MASC
reports exactly that of the detectors it was measuring
(Section~\ref{sec:related-neighbours}).
\end{enumerate}

Section~\ref{sec:scope} states exactly what is and is not claimed, and
Section~\ref{sec:threats} states what could make each claim wrong. The short version is in
the abstract and is worth repeating: this is $n=1$, self-measured, on a registry of my own
repositories.

\subsection*{A note on numbers and commits}

Line numbers and counts in a live repository are claims with a short shelf life, and an
earlier draft of this paper went stale by citing two of them. Every file and line citation
below is therefore pinned to a commit. The pre-fix baseline is \texttt{f59fb62}; two
citations in Section~\ref{sec:fixes} resolve at \texttt{27809ce}; present-day figures are
measured at \texttt{92e4548}, the commit named in the Reproducibility and artifacts
section. That section is unnumbered, so it is named rather than cross-referenced.

\section{The system under study}
\label{sec:system}

\subsection{The instrument and the claim}

\vacname{} (Verifiable Agent Claims) is an evidence-bundle format plus an offline verifier.
A bundle pins every artifact by sha256, declares headline numbers in
\texttt{results.summary}, and declares \texttt{results.checks} that bind those numbers to
recomputable quantities inside the artifacts. On success the verifier prints:

\begin{Verbatim}[frame=single,fontsize=\small]
structural verification: PASS (valid)
  proved offline: manifest schema, artifact presence + sha256, bundle closure,
  stated limitations, stamp agreement, declared results recomputed from artifacts.
\end{Verbatim}

That last clause is the load-bearing one. Everything in this paper is an attack on it.

Accepted bundles are copied into a public registry (\texttt{vac/registry.py:171} copies
\texttt{results.summary} verbatim) and rendered on a published page
(\texttt{index.html:113}). Both citations resolve unchanged at \texttt{f59fb62} and at
\texttt{92e4548}. A lie that survives the verifier is not academic; it is published.

The trust boundary is therefore between an issuer, who supplies the bundle, and a reader,
who sees the registry. The verifier is the only thing in between. In the vocabulary of
RFC 9334 \citep{birkholz2023rats}, the verifier appraises evidence against an appraisal
policy and produces a result that a relying party consumes. What that architecture does not
specify, and what this paper is about, is how anyone establishes that the appraisal policy
was actually applied.

\subsection{The external audit and the hardening baseline}
\label{sec:audit}

Giulio D'Erme, author of \texttt{cca-audit}
(\url{https://github.com/GiulioDER/cca-audit}), ran a deep audit after I posted an open
replay request. Four pull requests came out of it. Two are analysed here, \#1 and \#2; the
other two arrived after the measurements in this paper were taken and are recorded at the
end of this section.

His central finding: a bundle declaring \texttt{summary.verdicts: 9999} while the artifact
held 3, with \texttt{evidence/bundle.json} replaced by the four bytes \texttt{null} and its
sha256 re-pinned honestly. It exited 0 and printed \texttt{structural verification: PASS}.
The hash binding was intact. The recomputation ran over nothing, found no mismatch, and
passed. This is the textbook vacuity example transposed into a verifier: every request is
eventually acknowledged, in an environment that never generates requests
\citep{simmonds2010vacuity}.

That first pull request opened with \textbf{eight} bundles that exited 0 under
\texttt{structural verification: PASS}. Five were false declared numbers, the class the table
above describes. The other three were content crossing the bundle boundary and are distinct
defects rather than variants: \texttt{\_safe\_relpath} accepting a drive-anchored path, which
is an arbitrary read and a content-confirmation oracle; \texttt{\_bundle\_root} counting only
directories, so a top-level sibling file rode along and the closure scan never saw it, which is
a closure hole of the same class as the symlinks in \#8 rather than traversal; and
\texttt{\_extract\_tar} accepting Windows-separator and root-anchored members, which is the one
that depends on the untested Python window. An earlier draft of this paper said five paths plus
path traversal, and attributed the untested-window dependency to the wrong one of the three. The
auditor corrected both on 18 August 2026. His second pull request found something worse in kind: no \texttt{encoding=} on
\texttt{read\_text} and \texttt{write\_text}, so the same bytes produced opposite verdicts
on a cp1252 host and a UTF-8 host. That is fatal for a protocol whose entire premise is that
a stranger gets my answer offline, and it is a bit-for-bit reproducibility failure of
exactly the kind the reproducible-builds literature exists to characterise
\citep{lamb2022reproducible,fourne2023flossing}.

Two things cut the other way and are worth recording. First, four of his strongest findings
were refuted by my own specification rather than by argument: the spec had already decided
those cases and said so. His words on that, posted publicly, were
``Documentation that refutes an auditor is rarer than the bugs, and I would rate it higher''
\citep{derme2026devto}. He also names the four: unknown severity weighing 0, a score of 1.0
when nothing applied, the floor inequality, and the registry HEAD question. A separate sentence
in the pull request body makes the same point: ``four findings refuted by documentation rather
than by argument is not the usual experience reading someone's spec'' \citep{derme2026pr1}.

A draft of this paper carried a footnote asserting that the first of those two
sentences was my own compression and that he had not written it. That footnote was wrong, and
wrong in the direction that matters: it credited me with his sentence. It was introduced by a
pre-publication audit that checked the pull request body, found the second sentence there, and
inferred the first was fabricated. He wrote both, in different places, and corrected the record
himself. The episode is left visible here because a paper about checks that pass without
checking should not hide the one instance it committed against a named third party, and because
the failure has a shape worth naming: a correction is a new claim, and it needs the same
evidence the original did.

Second, my \texttt{RESULTS.md} byte-identity check caught mojibake in his own output
mid-review. The instrument worked on the auditor. That is checkable too: he reports it in the
same public comment, saying he fetched the source with a locale decode, mojibaked every section
sign and dash in the file, and that the byte-identity check caught it immediately
\citep{derme2026devto}. An earlier draft called this a first-person report resting on no public
artifact. It rests on the same artifact as the quote.

Everything in Section~\ref{sec:forgeries} runs against \texttt{main} at \texttt{f59fb62},
\emph{after} that soundness fix. That is what makes the four findings interesting rather
than routine: they survived an external audit and its remediation.

The audit did not stop with those two pull requests, and this record should not either. On
18 August 2026, after the hardening described above and after every present-day figure in
this paper was measured, the same auditor filed two more. Pull request \#9
(\url{https://github.com/egnaro9/vac-protocol/pull/9}) reports that a default Git for
Windows clone rewrites the repository's text artifacts to CRLF, so the sha256 pins in an
honest bundle stop matching; it changes only \texttt{.gitattributes} and a test, and it is
merged at \texttt{81f50cf}, after the commit the figures below are pinned to. Pull request
\#8 (\url{https://github.com/egnaro9/vac-protocol/pull/8}) is the one that belongs to this
paper's class, and it is open.

Its second case, E2 in that pull request's own table of reproductions, is a fifth vacuous
pass against the hardened verifier: the four of
Section~\ref{sec:forgeries} plus this one, found after them. It is deliberately not folded
into the five classes this paper claims (Section~\ref{sec:scope}, item 1), which are those
four plus the auditor's \texttt{null}-artifact case against the pre-hardening version.
Replace a covered artifact with a symlink whose target sits outside the bundle, and the
verifier exits 0 and prints \texttt{structural verification: PASS}, banner clause
\texttt{bundle closure} included. The bytes hash identically because they are the same
bytes; they simply live somewhere else on the verifying host. I reproduced it at
\texttt{92e4548}, the commit this paper's present-day figures are measured at, with controls
in both directions: an unlisted file placed directly in the bundle is refused
(\texttt{unlisted-file}), one byte appended to a covered artifact is refused
(\texttt{sha256-mismatch}), the symlinked artifact passes, and removing the out-of-bundle
target turns the same bundle into \texttt{missing-artifact}. That last control is why it is
recorded as a vacuous pass and not as a fifth forgery: it is host-dependent, and a stranger
replaying the bundle without the target file present gets a refusal, so the four of the
title stand. The mechanism is one line, \texttt{vac/verify.py:224} at \texttt{92e4548}: the
closure scan is \texttt{rglob} filtered by \texttt{is\_file()}, and \texttt{rglob} does not
descend a symlinked directory while \texttt{is\_file()} follows a symlinked file. The pull
request's first case is the other half of that same line: an unlisted file placed inside a
symlinked subdirectory is never seen at all, and the run exits 0 naming nothing, where the
same file placed directly in the bundle is refused.

That pull request is open and is deliberately not landed before this paper. Its own commit
adds two refusal statements to \texttt{vac/verify.py}, which moves the population every
number in Section~\ref{sec:results} is scored against: 146 raw refusal sites at
\texttt{92e4548}, 153 at the pull request's parent, which is \#2, and 155 at its own head.
Landing
it would invalidate the figures rather than update them, so it is recorded here as an open
finding against the fixed verifier and left for the next measurement round. A check that
prints \texttt{bundle closure} while reading evidence from outside the bundle is the same
shape as everything in Section~\ref{sec:forgeries}, which is the point worth stating
plainly: the class kept being findable after I started writing about it, and the person who
found it again was the auditor, not the author.

\section{The class: a check that cannot fail}
\label{sec:class}

\subsection{Definition}

Every finding in this paper is an instance of one shape.

\begin{quote}
A gate reports success along a path where it never examined the thing it claims to examine.
The pass is real. The checking never happened.
\end{quote}

\begin{definition}[Vacuous pass]
A verifier exhibits a \emph{vacuous pass} when it returns acceptance while an intended
binding, comparison, or coverage obligation was absent, bypassed, or semantically
unexamined.
\end{definition}

The common ancestor is an absence-assertion with no liveness proof. ``Nothing bad was
found'' is meaningful only if you have separately established that the detector \emph{can}
fire.

The weakness taxonomy already has vocabulary for the parts. CWE-703 is the pillar,
improper check or handling of exceptional conditions \citep{cwe703}; CWE-754 covers the
plain unchecked case and, by its own scope note, is not restricted to exception handling
\citep{cwe754}; CWE-636 names the design decision to fall back to a less secure state, with
the alternate term ``failing open'' \citep{cwe636}; CWE-390 is the tightest fit for the
instrument failures in Section~\ref{sec:instruments}, where an error condition was detected
and converted into a success signal \citep{cwe390}. Section~\ref{sec:sec-forge-summary}
concerns consistency between fields of one complex input, which is CWE-1288
\citep{cwe1288}. Having the names is a convenience, not a result. It does mean the class is
standard rather than idiosyncratic.

\subsection{The same shape elsewhere}

I have hit this shape repeatedly in work that had nothing to do with this protocol. These
are the author's own observations, offered as motivation rather than as data:

\begin{itemize}
\item a determinism test that compared two runs of the \emph{same} runtime stayed green
through a 69-diff cross-runtime divergence;
\item a health endpoint returned 200 while the real code path was dead;
\item a shell gate of the form \texttt{... \textbar{} grep "Tests "} matched the tally line whether
the tests passed or failed;
\item a test that sampled an \emph{animating} value passed against the very bug it targeted.
\end{itemize}

The published record contains the industrial-scale version. Certificate validation was
broken across payment SDKs, e-commerce platforms, and web-services middleware, with badly
designed APIs diagnosed as the root cause \citep{georgiev2012dangerous}. A production
attestation verifier, \texttt{cosign}, reported a false positive for
\texttt{verify-attestation -{}-type} when at least one attestation carried a valid
signature \emph{and} no attestation of the requested type was present: CVE-2022-35929
\citep{cve2022sigstore}. Both conditions are necessary, and the advisory states them as a
conjunction; the second is the one that makes it this paper's shape. That last one deserves
emphasis, because it is the closest external corroboration this paper has. It is the same
shape as Section~\ref{sec:sec-forge-stamp}, in somebody else's verifier, with a CVE
attached. The mechanisms differ: cosign applies a type filter to the wrong subject, while
Section~\ref{sec:sec-forge-stamp} skips a comparison guarded on operand existence. What is
shared is an obligation over an absent subject discharged as success, with the party
supplying the input deciding the absence.

\section{Related work}
\label{sec:related}

The draft of this paper that circulated privately said only that ``the territory is already
occupied'' and named no occupants. This section names them, and in three places the
literature narrows what I can claim.

\subsection{Vacuity and sanity checking}

Vacuity detection asks whether a property was satisfied trivially. Beer et al.\ introduced
it after observing that implications can be satisfied by antecedent failure, and reported
from IBM hardware-verification practice that during first formal verification runs
``typically 20\% of formulas are found to be trivially valid, and that trivial validity
always points to a real problem in either the design or its specification or environment''
\citep{beer2001vacuity}. Kupferman and Vardi extended subformula-replacement vacuity from
the ACTL fragment to CTL* \citep{kupferman2003vacuity}. Kupferman's survey frames the whole
family as \emph{sanity checks}, motivated by the observation that ``when the answer to the
correctness query is positive, most model-checking tools provide no additional
information'', and argues that vacuity and coverage ``are essentially the same: both are
based on repeating the verification process on some mutant input''
\citep{kupferman2006sanity}. Simmonds et al.\ supply the canonical illustration: a property
that every request is eventually acknowledged is satisfied in an environment that never
generates requests \citep{simmonds2010vacuity}.

The shape is the same as mine. The mechanism is not, and I want to be exact about the
difference rather than borrow the formalism. Vacuity in model checking is a property of a
specification relative to a model, decided by replacing subformulas. A vacuous pass here is
a verifier whose code path returns acceptance without executing a comparison. This paper
does not instantiate Beer et al.'s formal definition, and it should not be read as claiming
to. What it borrows is the framing: a positive answer is the thing that needs a further
automatic check. The 20\% figure is also the closest published precedent for the shape of my
own baseline result, and it is an independent industrial measurement rather than a
self-report.

\subsection{The oracle problem and oracle adequacy}

Weyuker named the assumption that a tester or external mechanism can decide whether output
is correct, and the programs for which it fails \citep{weyuker1982nontestable}. Barr et al.\
survey the resulting problem and give the formalism this paper needs: the ground truth is a
conceptual oracle that always gives the right answer, unknowable in all but trivial cases,
against which real oracles are sound or complete only in part \citep{barr2015oracle}. Their
category of \emph{implicit} oracles, built on general anomaly detection such as abnormal
termination, names exactly what several of my instruments were doing while claiming to be
specified oracles: a sweep that scored ``nonzero exit'' as ``correctly refused''
(Section~\ref{sec:instruments}) was running on an implicit oracle and reporting a specified
one.

Oracle quality has been measured directly. Schuler and Zeller's checked coverage measures
the fraction of executed statements that actually influence an oracle, motivated by a parser
with 83\% statement coverage in which ``none of the parsed results is actually checked for
any property'' \citep{schuler2011checkedcov,schuler2013checkedcov}. Their internal-validity
paragraph carries the warning this paper should read against itself: ``even programs without
oracles can still achieve a high mutation score by relying on uncontrolled implicit
checks''. Huo and Clause measure the same phenomenon from the other side, detecting brittle
assertions and \emph{unused inputs}, inputs provided by a test that no assertion checks
\citep{huo2014oracle}; that is the closest published analogue to the closure rule in
Section~\ref{sec:sec-forge-check}, where an evidence artifact covered by no check becomes a
named refusal. Jahangirova et al.\ use mutation to expose oracle false negatives and then
\emph{improve} the oracle \citep{jahangirova2016oracle}, which is methodologically what
Section~\ref{sec:liveness-tests} does: one test per surviving refusal, so the gate is
asserted rather than merely executed.

On coverage as a proxy, Inozemtseva and Holmes generated 31,000 suites over five systems and
found only a low to moderate correlation between coverage and effectiveness once suite size
is controlled, concluding that coverage ``should not be used as a quality target''
\citep{inozemtseva2014coverage}. Their effectiveness measure is itself the mutation score,
which is worth stating plainly rather than citing them as evidence that mutation works.
Zhang and Mesbah composed 6,700 suites from 24,000 assertions and found assertion count and
assertion coverage strongly correlated with effectiveness \citep{zhang2015assertions}; their
introduction states the thesis of my Section~\ref{sec:liveness-tests} directly, that
coverage without checking for correctness is meaningless.

\subsection{Mutation testing and statement deletion}

Mutation analysis originates with Hamlet \citep{hamlet1977compiler} and DeMillo, Lipton and
Sayward \citep{demillo1978hints}. Budd et al.\ treat mutation as an \emph{adequacy}
criterion on test data rather than a correctness argument \citep{budd1980theoretical}, and
Zhu, Hall and May's survey establishes adequacy as the formal category a mutation score
belongs to \citep{zhu1997adequacy}. That distinction is what licenses the careful reading of
my own 1.000 in Section~\ref{sec:limits-score}. The surveys by Jia and Harman
\citep{jia2011mutation} and Papadakis et al.\ \citep{papadakis2019mutation} are the
reference points for the method as an experimental discipline. Just et al.\ supply the
validity argument and its ceiling: across 357 real faults in five applications, mutant
detection correlates with real fault detection independently of coverage, 73\% of real
faults are coupled to mutants generated by common operators, and 17\% are coupled to no
mutant at all \citep{just2014mutants}. Their finding that statement deletion is one of the
three operators most often coupled to real faults is what turns my choice of a deletion
operator from an ad hoc script into a defensible one. Petrovi\'c and Ivankovi\'c report
mutation at Google, including the practice of excluding ``arid'' lines explicitly rather
than counting them as caught \citep{petrovic2018google}, which is the industrial precedent
for the exclusion set in Section~\ref{sec:exclusions}.

The operator itself is not new, and this is the first place the literature narrows my claim.
Untch proposed statement deletion (SDL) as a single sufficient mutagenic operator
\citep{untch2009sdl}, and Deng, Offutt and Li evaluated it empirically
\citep{deng2013sdl}. Deleting error-handling specifically is also published: Ji et al.\
define a catch-block deletion operator \citep{ji2009exception}, and Kumar et al.\ define
throw-statement deletion, which removes the sole statement by which a program signals
rejection and asks whether any test fails \citep{kumar2011mutants}. That is structurally
identical to replacing a refusal statement with \texttt{pass}. Loise et al.\ introduce
fifteen security-aware operators for Java and report that standard operators are unlikely to
introduce comparable vulnerabilities \citep{loise2017security}, which is the published
precedent for my claim that a general-purpose mutation run would not have targeted the
refusal sites.

Extreme mutation is the second narrowing. Niedermayr et al.\ delete whole method bodies to
find \emph{pseudo-tested} methods, those tested such that faults would not be detected
\citep{niedermayr2016extreme}, and Vera-P\'erez et al.\ study the phenomenon at scale and
report that developers respond by strengthening tests rather than deleting code
\citep{veraperez2019pseudo}. An earlier draft of this paper claimed statement granularity as
the residual novelty over that work. That is wrong: Maton, Kapfhammer and McMinn published
exactly it, using SDL to uncover pseudo-tested \emph{statements}, finding 722 cases across
27 Java projects of which 48\% lie outside pseudo-tested methods
\citep{maton2024pseudotested}. Their paper even makes my own argument about tool blind spots,
that a popular Java mutation tool would not have mutated some of the statement types
involved.

\subsection{Mutation pointed at checkers rather than at code under test}

A smaller literature points mutation at the checking apparatus. Martin and Xie mutate
XACML access-control \emph{policies} \citep{martin2007acpolicy}. Ami et al.'s muSE uses
mutation to discover unsound choices in security-focused Android static analysis
\citep{ami2021muse}, and their MASC carries the same idea into crypto-API misuse detection
at a scale of 20,303 mutants against nine detectors \citep{ami2022masc,ami2023masc}.
G\"orz et al.\ extend mutation analysis to grading fuzzers
\citep{gorz2023fuzzers}. Chen and Furia perform robustness testing of intermediate
verifiers, generating semantically equivalent Boogie programs and finding brittle behaviour
in 16 of 135 \citep{chen2018robustness}. Lipp et al.\ supply the external calibration a
reader should apply to my 0.330: state-of-the-art static C analyzers miss between 47\% and
80\% of vulnerabilities in a benchmark of real programs \citep{lipp2022static}.

The direction matters and I state it rather than leave a reviewer to find it. muSE, MASC,
G\"orz et al., Delcourt et al.\ and Chen and Furia all mutate the checker's \emph{input},
seeding defects into the artifact under analysis and asking whether the checker detects
them. Chen and Furia
additionally target the opposite failure from mine, false alarms rather than false passes.
This paper mutates the checker's own refusal statements. Similarly, the existing
mutation-in-continuous-integration literature puts the mutation tool \emph{inside} the
pipeline: \"Org\aa rd et al.'s industrial case study evaluates C++ mutation tools in GitHub
Actions workflows and says nothing about mutating a gate \citep{orgard2023mutationci}. This
paper points the mutation tool \emph{at} the gate. That is a different move from grading a
checker by mutating what it reads, which is muSE's, MASC's and, concurrently, Delcourt et
al.'s; Section~\ref{sec:related-neighbours} sets out how much of the surrounding ground
those three occupy.

\subsection{Three close neighbours}
\label{sec:related-neighbours}

Three papers sit close enough to this one to change what I am entitled to claim. One is
prior art I should have found earlier: MASC, published at IEEE S\&P in 2022. The other two
were posted to arXiv while this study was running, on 21 June and 14 August 2026, so they
are contemporaneous work rather than literature I failed to search. Neither of those two is
in print; one is a preprint submitted to a magazine, the other is accepted at a conference
that meets in October 2026. I describe each by what it already owns.

\paragraph{MASC.} Ami et al.\ carry muSE's idea into crypto-API misuse detection
\citep{ami2022masc,ami2023masc}. Twelve usage-based mutation operators, three mutation
scopes, and an explicit three-adversary threat model instantiate compilable variants of
crypto-API misuse cases, seeded into thirteen Android apps and four Apache Qpid Broker-J
subsystems: 20,303 mutants, generated in about fifteen minutes. Nine crypto-detectors were
evaluated and 19 unique undocumented flaws found, appearing across those detectors as 76
flaw instances, of which 45 (59.21\%) are attributable solely to the mutation approach
rather than to base instantiations of the same misuse cases; all flaws in six of the nine
detectors were visible only through MASC. The journal extension enlarges this to nineteen
operators, 107 misuse cases and five further detectors. MASC belongs in the list above, and
its absence from an earlier draft of this paper was an oversight rather than a judgement,
particularly since muSE is already cited here. Three of its results bear on mine directly,
and a fourth, the base-instantiation comparison of its Step 6, is taken up in
Section~\ref{sec:related-unclaimed}.

First, its motivating premise is that hand-curated benchmarks are, in their words,
``incomplete, incorrect, and impractical to maintain'', hedged with a \emph{may be} and
evidenced by the OWASP benchmark treating AES in ECB mode as secure until 2020
\citep{ami2022masc}. That is the published form of the structural ceiling I derive in
Section~\ref{sec:rq1}, and the direction of the argument is theirs. The two claims do
different work, though. Theirs is empirical and hedged; mine is a bound, since a committed
corpus of $n$ fixtures cannot catch more than $n$ of 112 sites however well the $n$ are
written. The arithmetic is not a discovery. It is what makes the $10/112$ in
Section~\ref{sec:rq1} interpretable.

Second, their first operator is atypical letter case. Passed to
\texttt{Cipher.getInstance}, the algorithm name \texttt{"DES"} is flagged by CryptoGuard and
\texttt{"des"} is not. The forgery in Section~\ref{sec:sec-forge-severity} costs one capital
letter. I keep the
instance and withdraw any suggestion that the shape is unreported, because it has been in
print since 2022.

Third, and running the other way, before analysing uncaught mutants they found that some
detectors ``analyze only a limited portion of the target applications''. CryptoGuard and
CogniCrypt do not handle multiple dex files, which left CryptoGuard analysing 871 of 2,515
Android mutants until they patched it, and CryptoGuard skips any class whose package name
contains \texttt{android.} or ends in \texttt{android}, which excludes applications such as
LastPass and LinkedIn. They group these under a flaw class zero and exclude them from their
counts for a reason they state plainly: ``these gaps were not detected using MASC's
evaluation'' \citep{ami2022masc}. A detector returning a clean verdict over code it never
read is the shape this paper measures, and mutating the detector's input did not find it.
The mechanism is not identical: theirs is a scope limitation rather than a branch that
returns acceptance, and a silent detector reads as clean rather than issuing an explicit
verdict. With that caveat, it is the clearest published evidence I have that the two
mutation sites answer different questions.

\paragraph{Mutation testing of semantic judges.} Delcourt et al.\ point mutation at
LLM-based judges of domain class diagrams \citep{delcourt2026judge}. Eleven operators, five
rule-based and six LLM-assisted, inject semantic defects into PlantUML class diagrams while
leaving the paired textual description unchanged, and each operator also emits a short
natural-language description of the defect it injected. A mutant is killed when at least one
issue the judge reports reaches cosine similarity $\tau = 0.55$ against that description,
the threshold calibrated on 209 hand-labelled judgments to an F1 of 0.90. Across 547 mutants
and six judge configurations, 3,282 mutant judgments in total, the kill ratio is 82.4\%, and
their headline result is that this automated score largely reproduces the ordering a manual
precision study produces over the same configurations: Pearson $r = 0.624$, and the same
ordering on 11 of the 15 configuration pairs. They read per-operator kill ratios as fault
classes the judge misses, and they warn that easy operators inflate the aggregate. This
paper makes both of those moves, and neither of us invented them: reading survivors as fault
classes the oracle does not detect is textbook mutation testing, and their own background
section says so with a citation. What is worth stating plainly is that using a mutation tool
to score a grader, and reporting a kill ratio over it, is neither a thin nor an unoccupied
idea in 2026, and this paper does not present it as one.

What their eleven operators do not touch is the judge. Every one of them mutates the
artifact the judge is asked to judge. Judge variation in their design is configuration
selection, three models crossed with two prompts, which they describe as reflecting standard
practice in judge development; no operator is applied to a judge's source, to its prompt
text, or to the kill detector, and the similarity threshold is calibrated rather than
mutated. Two of their three models are closed-source, so for most of their configurations
there is no checking code available to mutate in the first place. Two further differences
matter for what each instrument can say. Their failure mode is judge sensitivity, whether a
model notices a defect it did in fact look at; a silent judge simply leaves the mutant alive
and pays for it in kill ratio. Mine is a path that returns acceptance without examining
anything, which has no analogue in their setting. And their kill decision is a calibrated
embedding threshold whose transferability across models they name as a threat to validity,
where the predicate in Section~\ref{sec:method} is exact and machine-decidable, which is the
ordinary situation for a mutation tool scored by a unit suite rather than an achievement of
this paper. That is a difference in what the two instruments can claim, not a ranking of the
two papers.

\paragraph{A green suite that keeps shipping defects.} Bilal and Mughal report a production
rental-search assistant whose suite reached 1,553 test cases in six weeks, passed
continuously, and kept shipping user-facing defects \citep{bilal2026allgreen}. They classify
all 252 bug-fix commits by the boundary, or seam, each defect escaped through, and find 110
of them, 44\%, in four seams no component-level test observes: the live browser runtime, the
non-default market, the end-to-end flow, and the whole-system level. Their mechanism is
substitution. Wherever a component test needs determinism, it replaces the side of the
boundary it cannot control with a fixed stand-in, and the seam then goes unobserved
entirely, so the count of such tests can rise without limit while the set of conditions the
suite ever meets stays where it was. That is the conclusion of Section~\ref{sec:whymissed}
reached in another domain, and reached two months before I reached it.

The mechanism differs, and the difference is worth keeping. Their blindness always sits at a
boundary where a stand-in took the place of something uncontrolled, and all four of their
seams are boundaries. The vacuous pass in Section~\ref{sec:vacuous1000} has no boundary in
it: the comparison runs to completion, inside one function, over an artifact with nothing in
it. What is shared is the conclusion, that adding tests inside the substituted world does
not move what gets observed.

One of their incidents is closer still. A quoting error in a client-side component broke a
form step, was visible only in a running browser, and shipped twice: the guard left behind by
the first fix lived in ``the same browser-blind harness that had missed the defect''
\citep{bilal2026allgreen}, so it could not see the repeat. That is an external instance of
the null result in Section~\ref{sec:rq2}, reported independently, on a system I did not
write. They also record that 107 of 252 fixes touched a test file, and treat that as an
upper bound on real regression guards, since a fix that edits a test file has not thereby
acquired a test that would catch its own defect again. That is the proxy the kill predicate
here exists to replace. Their evidence is broader than mine and shallower. It is
observational: they inject no faults, and by their own statement they count the defects they
found, so silent escapes are not measured. Nothing in their design can report what fraction
of their checks are inert. Nothing in mine generalises past one verifier.

\paragraph{What the three take, and what they leave.} Together they remove three things an
earlier draft of this paper was implicitly claiming. Using mutation to grade a grader is
occupied, at 20,303 mutants in one case and 3,282 judgments in another. Reading surviving
mutants as blind spots in the checker, and warning that easy mutants inflate the aggregate,
belongs to the mutation-testing literature, and to Delcourt et al.\ in this setting, before
it belongs to me. And a large continuously green suite that keeps shipping defects because
of what its checks observe rather than how many there are is Bilal and Mughal's reported
observation, not this paper's discovery. What none of the three does is modify a line of the
checker's own source: MASC and Delcourt et al.\ mutate the input, and Bilal and Mughal
inject nothing at all. None of the three reports its own measuring apparatus returning a
passing verdict along a path where it performed no comparison, which is the material in
Section~\ref{sec:instruments}, although MASC reports exactly that of the detectors it was
measuring. Section~\ref{sec:related-unclaimed} states what that leaves.

\subsection{Supply-chain attestation and verifier soundness}

The wider context is machine-readable evidence about software. TUF established that a
signature-valid update system can still fail once you look past the cryptography
\citep{samuel2010tuf}; Sigstore made signing and transparency broadly deployable
\citep{newman2022sigstore}; reproducible builds argue that trusting code is not the same as
trusting its executables \citep{lamb2022reproducible,fourne2023flossing}; Ladisa et al.\
taxonomise 107 attack vectors linked to 94 real incidents \citep{ladisa2023sok}; and Xia et
al.\ document practitioner experience with software bills of materials from 17 interviews
and 65 survey responses \citep{xia2023sbom}. Thompson's argument that inspecting a
checker's source tells you nothing if the tool that produced it is compromised is the
origin point for verifying the verifier \citep{thompson1984trusting}; what follows is the
mundane, non-adversarial version of the same regress.

Two entries in this literature bear directly on my Section~\ref{sec:sec-forge-check} rule,
and both cut against its novelty while supporting the paper's thesis. in-toto's layout
model already carries the closure obligation \citep{torresarias2019intoto}, and the
specification states it explicitly: there is an implicit \texttt{"ALLOW *"} at the end of
each rule list, and ``[n]ot including \texttt{"DISALLOW *"} at the end of a rule list could
allow artifacts to sneak into the step or inspection undetected'', so it ``is generally
recommended that all rule lists include a \texttt{"DISALLOW *"} at the end''
\citep{intoto2023spec}. A mature framework identified the same fail-open shape, wrote the
warning down, and still left the safe behaviour off by default. My rule is therefore a
re-derivation made mandatory, not a discovery. In the same direction, the in-toto attestation
Statement layer imposes no requirement that a verifier evaluate every subject
\citep{intoto2024attestation}, and the SLSA v1.2 build track states that ``SLSA v1.0 does not
have any requirements on the completeness or verification of \texttt{resolvedDependencies}''
while separately advising that ``[v]erification tools SHOULD reject unrecognized fields in
\texttt{externalParameters} to err on the side of caution'' \citep{slsa2025spec}. That second
rule is the fail-closed-on-unknown convention my Section~\ref{sec:sec-forge-severity} defect
violates, one scope over.

\subsection{Evaluation harnesses and leaderboards}

The bundles this verifier accepts are, in practice, evaluation results. Biderman et al.\
name sensitivity to evaluation setup and lack of reproducibility as structural problems of
language-model evaluation, written from years of operating a harness
\citep{biderman2024lessons}. Zhao et al.\ study 1,045 foundation-model leaderboards and
catalogue recurring workflow smells \citep{zhao2025lbops}. Together these are the demand-side
argument for an offline recomputable evidence bundle, and the reason the host-dependence
result in Section~\ref{sec:host} belongs to a studied class rather than to personal
carelessness.

\subsection{What is left unclaimed}
\label{sec:related-unclaimed}

After all of that, the honest residue is narrower than the earlier draft implied. Statement
deletion is Untch's. Extreme mutation is Niedermayr's, and its statement-granularity form is
Maton's. Deleting the statement that signals rejection is Kumar's. Mutating a checker to
grade it is muSE's, MASC's and Chen and Furia's, and as of August 2026 it is also Delcourt
et al.'s in the semantic-judge setting \citep{delcourt2026judge}. Reporting that a large
continuously green suite keeps shipping defects because of what its checks observe is Bilal
and Mughal's \citep{bilal2026allgreen}. The green-baseline precondition in
Section~\ref{sec:gate} is long-standing default behaviour in PIT, a widely used Java
mutation engine, which validates baseline suite state before running
\citep{pitest_faq}; my version of that rule is a rediscovery of standard tooling practice,
and saying so costs nothing.

What remains unclaimed by prior work, as far as I can establish, is the combination of two
things. The \emph{site selection}: refusal statements enumerated as a population, each the
sole point of rejection for one class of bad input, so that the denominator of the score is
the checker's own refusals rather than an arbitrary set of program points. And the
\emph{subject}: not the artifact the checker reads, which is what MASC and Delcourt et al.\
mutate, but the checker's own acceptance logic, whose verdict is published. Each of the two
is a specialisation of something already published, and the combination is narrow.

The two-detector kill predicate is not a third novelty, and an earlier draft of this section
implied that it was. Deciding a kill by exact, machine-checkable suite state is the ordinary
arrangement in conventional mutation testing; it looks distinctive here only against two
comparators that could not have had it, one grading language models by embedding similarity
and the other triaging its unkilled mutants by hand. What the predicate buys is not novelty
but decomposition. Because the corpus disjunct can be scored with the unit-suite disjunct
removed, the corpus written to prove the verifier can refuse can be asked what it catches on
its own, which is 10 of 112 (Section~\ref{sec:rq1}), the paper's least comfortable number.
MASC reports the same shape of result from the other side of the pipeline, attributing 45 of
76 crypto-detector flaws to mutation and only 31 of 76 to literal base instantiations of the
same misuse cases \citep{ami2022masc}. The partition differs: theirs is between two stimulus
sets against one detector, mine between two detectors over one mutant population, so my
number scores a standing regression asset rather than a baseline for the mutants. The
direction of the finding is the same in both, and it was theirs first.

Read strictly, then, this is a case study carrying a specialised instrument rather than a new
method, and the contributions in Section~\ref{sec:intro} are worded to say so.

\section{Four forgeries against the hardened verifier}
\label{sec:forgeries}

All four run against \texttt{main} at \texttt{f59fb62}, after the soundness fix of
Section~\ref{sec:audit}. Every run is preceded by a liveness control:
\texttt{python -m vac.verify fixtures/valid} must exit 0. Without that control a sweep proves
nothing, and Section~\ref{sec:instruments} records exactly that going wrong twice.

\subsection{The lie typed as a string}
\label{sec:sec-forge-summary}

\texttt{results.summary} values are compared against recomputed quantities by a walk that
returns early on any leaf that is not \texttt{int} or \texttt{float}. Retype every summary
number as a JSON string and nothing is ever compared. No artifact is touched. No hash is
re-pinned.

\begin{Verbatim}[frame=single,fontsize=\small]
summary: {"verdicts": 3, "fixed": 2, ...}
     ->  {"verdicts": "9999", "fixed": "9999", ...}
exit=0   structural verification: PASS
\end{Verbatim}

The specification blessed this. \texttt{SPEC.md:139} at \texttt{f59fb62}, inside section 2.5
\texttt{results}, reads in full: ``Non-numeric, descriptive values pass through.'' This is
therefore a specification hole, not an implementation slip. The spec permitted a class of lie
it had not imagined. It is also, in taxonomy terms, a failure to validate consistency between
elements of one complex input \citep{cwe1288}: the headline field and the artifact it
summarises must agree, and retyping the field removes the comparison rather than failing it.

\subsection{Deleting the check instead of breaking it}
\label{sec:sec-forge-check}

\texttt{\_validate\_manifest} requires only that \texttt{results.checks} be a non-empty list.
Nothing requires a listed evidence artifact to be covered by any check. Delete the check that
recomputes a number and the number moves from the strictly-bound branch to a loose branch
that accepts any value appearing anywhere in the recomputed pool.

\begin{Verbatim}[frame=single,fontsize=\small]
checks: [certlab-bundle-v1, fleet-board-v1, evalmut-run-v1,
         crashkit-battery-v1, modeldrift-board-v1]
    ->  [fleet-board-v1, evalmut-run-v1, crashkit-battery-v1,
         modeldrift-board-v1]
exit=0   structural verification: PASS
\end{Verbatim}

The audit's fix made \emph{breaking} a check a named refusal. It left \emph{deleting} one
free. This is the general lesson of hardening: constraining how a check fails does not
constrain whether it runs. As Section~\ref{sec:related} notes, in-toto's specification
already recommends the catch-all rule that closes this, as an opt-in
\citep{intoto2023spec}; the claim here is about my own specification, not about attestation
frameworks in general.

\subsection{One capital letter}
\label{sec:sec-forge-severity}

The crashkit profile weights findings by severity. At \texttt{f59fb62}:

\begin{Verbatim}[frame=single,fontsize=\small]
_CRASHKIT_WEIGHTS = {"none": 0, "low": 1, "med": 2, "high": 4,
                     "critical": 8}                          # verify.py:548
total_w  = sum(_CRASHKIT_WEIGHTS.get(c.get("severity"), 0) ...)  # verify.py:594
failed_w = sum(_CRASHKIT_WEIGHTS.get(c.get("severity"), 0) ...)  # verify.py:596
\end{Verbatim}

\texttt{.get(sev, 0)} maps any unrecognised label to zero. Re-case the severity on the
\emph{failed} cases only and their weight vanishes from the numerator while the passing cases
keep theirs. The score divides to \texttt{0.0} through the ordinary arithmetic path. It never
touches a guard, an exception, or a fallback branch.

\begin{Verbatim}[frame=single,fontsize=\small]
case severities   [critical, critical, high, med, med]
   graded subset  [critical, critical, high, med]   (one med case is
                                                     truncated, so it is
                                                     excluded from grading)
   honest weights total 8+8+4+2 = 22, failed 8+2 = 10
   vulnerability_score = 10/22 = 0.4545

tampered           [critical, Critical, high, Med, Med]  (3 re-cased,
                                                          failed rows only)
   graded weights  total 8+0+4+0 = 12, failed 0
   vulnerability_score declared 0.0
   evidence/eval_run.json sha256 re-pinned HONESTLY
exit=0   structural verification: PASS
\end{Verbatim}

The truncated-case exclusion is spelled out because without it the arithmetic is not
reproducible from the printed severity list, and a paper about offline recheckability cannot
afford an example a reader cannot recompute.

At \texttt{f59fb62}, \texttt{grep -rn severity tests/} returned zero matches across all three
test files. And \texttt{SPEC.md:306-307} at the same commit stated the rule plainly, that the
severity weights are ``the same weights crashkit grades with; an unknown severity weighs 0'',
so, as in Section~\ref{sec:sec-forge-summary}, the specification wrote the hole down and
called it a decision. That is the CWE-636 pattern precisely: a design that falls back to a
more permissive state \citep{cwe636}. At \texttt{92e4548} the same grep returns 13 matches,
because the fix brought tests with it.

A security score of \texttt{0.0} on a bundle whose own rows record three failed cases is then
copied verbatim into the public registry.

\subsection{Deleting the stamp rather than faking it}
\label{sec:sec-forge-stamp}

Four stamp comparisons are guarded on the artifact-side key existing
(\texttt{verify.py:284, 288, 374, 383} at \texttt{f59fb62}). An issuer who deletes those keys
and re-pins honestly passes all four without a single comparison executing, while the report
still claims \emph{stamp agreement}.

This one carries the cleanest discriminator in the paper, because the same fields corrupted
rather than removed are refused by name:

\begin{Verbatim}[frame=single,fontsize=\small]
WRONG VALUES (harness_commit/taskset_hash/prompt_hash = "DEADBEEF")
    exit=1   FAIL stamp-mismatch: taskset_hash: protocol 00112233445566aa,
                                                artifact DEADBEEF
             FAIL stamp-mismatch: prompt_hash:  protocol aabbccdd00112233,
                                                artifact DEADBEEF
             FAIL stamp-mismatch: harness_commit: protocol f1e2d3c,
                                                  artifact DEADBEEF

KEYS DELETED (same fields, removed; sha256 re-pinned honestly)
    exit=0   structural verification: PASS
             protocol.hashes still declares all 7 pins
\end{Verbatim}

The check is demonstrably alive and fires on corruption. It is blind to absence. Another
profile fails closed on exactly this case one branch along: \texttt{verify.py:642} at
\texttt{f59fb62} compares the pinned hash against the artifact's \texttt{git\_sha} directly, so
a deleted key compares unequal and is refused by name rather than skipped. The behaviour is
therefore an asymmetry rather than a design decision. Two nearby comparisons that look like
counterexamples are not: \texttt{verify.py:639} guards the \emph{protocol} operand, firing when
the key is absent from \texttt{protocol.hashes}, which is the reverse of this forgery, and the
modeldrift stamp check at \texttt{verify.py:1000} compares against a recomputed fingerprint, so
its artifact operand cannot be deleted at all. It is
CWE-754 in its adversarial reading: the programmer assumed a condition would not occur, and
the party who decides whether it occurs is the issuer \citep{cwe754}.

The general form is worth stating: \textbf{a comparison guarded on both operands existing is
not a check, it is a suggestion.} The party supplying one operand decides whether the
comparison happens.

\section{Method: refusal-site deletion}
\label{sec:method}

The four findings above were found by hand, which makes them anecdotes. To turn them into a
measurement I applied a mutation operator to the verifier itself, implemented as
\texttt{tools/mutation\_sweep.py} in the repository.

\subsection{The operator}

\begin{definition}[Refusal site]
A \emph{refusal site} is one source statement that appends a named failure reason to the
failure list returned by \texttt{verify\_bundle}. Operationally the sweep selects lines
matching the regular expression
\begin{Verbatim}[fontsize=\small]
^\s*(f|failures)\.append\(
\end{Verbatim}
in \texttt{vac/verify.py}. In that file the live idiom is \texttt{f.append(}; the
alternation exists because \texttt{vac/registry.py} uses the longer name. An earlier draft
of this paper described the sites as \texttt{failures.append} statements, which was simply
wrong for the file being measured; the wrong name had been copied from the sweep's own
docstring.
\end{definition}

\begin{definition}[Operator]
The \emph{refusal-site deletion} operator replaces one refusal site with \texttt{pass},
generating one mutant per site. This is statement deletion \citep{untch2009sdl,deng2013sdl}
restricted to a chosen statement class, in the same family as throw-statement deletion
\citep{kumar2011mutants} and catch-block deletion \citep{ji2009exception}.
\end{definition}

At \texttt{f59fb62} the file contains 112 refusal sites. At \texttt{92e4548} it contains 146,
of which 3 are excluded (Section~\ref{sec:exclusions}), leaving a denominator of 143. Both
counts are reproducible with the grep above.

\subsection{The kill predicate}

\begin{definition}[Caught]
A mutant is \emph{caught} if any of the following holds: (i) the unit suite fails; (ii) the
liveness control breaks, that is \texttt{python -m vac.verify fixtures/valid} stops exiting
0; or (iii) any committed tamper fixture stops being refused, that is stops exiting 1.
Otherwise it \emph{survives}.
\end{definition}

The three-way disjunction is deliberate, and so is the order. The sweep evaluates the
disjuncts in the order written and reports the first that fires, which makes the score cheap
but makes the per-disjunct attribution \emph{marginal} rather than absolute: the fixture
disjunct is only ever reached on mutants the unit suite has already missed. That asymmetry
matters enough that Section~\ref{sec:rq1} scores the fixture corpus a second time as a
detector standing on its own, and the two numbers are not the same.

\begin{definition}[Score]
$\mathrm{score} = \dfrac{\text{mutants caught}}{\text{refusal sites} - \text{exclusions}}$.
\end{definition}

This is an adequacy measurement in the sense of \citet{budd1980theoretical} and
\citet{zhu1997adequacy}: it measures whether a suite detects a chosen class of artificial
change, never whether the verifier is correct.

\subsection{Exclusions}
\label{sec:exclusions}

Some refusals cannot be reached by any bundle-shaped input. Counting them as caught would
inflate the score; leaving them in as permanent survivors would make the denominator lie.
The sweep therefore carries an explicit exclusion set, keyed by a distinctive source fragment
together with the number of lines that fragment is \emph{expected} to match, so that a
rename or a refactor aborts the run rather than silently resizing the denominator in either
direction. This mirrors the arid-line practice reported at Google
\citep{petrovic2018google}.

At \texttt{92e4548} the exclusion set holds two fragments covering three source lines, and
both are \texttt{OSError} wrappers:

\begin{itemize}
\item one line wrapping the read of \texttt{RESULTS.md}: to reach the check the artifact must
already be in the trusted set, which required an \texttt{is\_file()} plus a full sha256 read
of the same bytes, so only a filesystem race between the hash pass and this read could fire
it;
\item two lines wrapping reads in both render comparators, for the same reason: a render not
listed in evidence is refused earlier as \texttt{check-artifact-not-listed}, and one that is
listed was already opened and hashed before the check runs.
\end{itemize}

Both were probed empirically rather than assumed. An earlier draft described the two
exclusions differently, as one unreachable \texttt{flips.json} branch and one \texttt{OSError}
wrapper; that description matched an intermediate state of the code and no longer matches the
exclusion set, which is why the composition is restated here rather than referenced.

\subsection{The validity gate}
\label{sec:gate}

The sweep aborts unless the unmutated baseline is green. This is a precondition of the
measurement, not a convenience, and Section~\ref{sec:vacuous1000} is the incident that
bought it. It is also standard: PIT validates baseline suite state before running mutation
analysis \citep{pitest_faq} and refuses outright when it is red. That refusal is not on the
cited FAQ page, which lists only causes of a green-locally, red-under-PIT discrepancy; it is
in the source, where \texttt{DefaultCoverageGenerator} throws
\texttt{PitHelpError(Help.FAILING\_TESTS)}, whose message reads ``Mutation testing requires a
green suite'', with \texttt{skipFailingTests} defaulting to \texttt{false}.

The gate covers two things and not a third, and the gap is worth naming because it is the
same shape as everything else here. It covers baseline greenness, and it covers exclusion
arity: each excluded fragment declares how many lines it should match, and a mismatch aborts.
It does not cover the \emph{site population}, which is the denominator. Nothing asserts how
many refusal sites the enumeration should find, so a behaviour-preserving refactor can resize
the denominator silently. Demonstrated on a clean copy of \texttt{92e4548}: moving four
\texttt{stamp-mismatch} refusals into a one-line \texttt{\_refuse(f, msg)} helper leaves the
suite byte-identical (237 passed, 16 skipped, 4 xfailed, plus one pre-existing failure on the
interpreter used, both before and after), keeps both exclusion counts matching so nothing
aborts, and moves the enumeration from 146 raw sites to 143, a denominator of 140 rather than
143. Four distinct refusal obligations now share one site whose deletion breaks all four, so
that site is trivially killed: real coverage falls while the score rises. CI's
\texttt{-{}-floor 0.99} is a floor on the ratio and would not notice. Pinning an expected site
count the way \texttt{EXCLUDE} pins its arity is the obvious fix. This paragraph describes
\texttt{92e4548}, where it was not done; the pin landed afterwards at \texttt{e932868}, and run
against that revision the same refactor aborts the sweep with exit 2, reporting 143 raw sites
and 140 scored where 146 and 143 are declared, instead of scoring the smaller denominator
(Section~\ref{sec:fixes-open}).

\section{Results}
\label{sec:results}

\subsection{RQ1: baseline refusal-site liveness}
\label{sec:rq1}

At \texttt{f59fb62}, with a 114-test suite (measured by \texttt{pytest -{}-collect-only}) and
16 committed tamper fixtures:

\begin{Verbatim}[frame=single,fontsize=\small]
112 mutants, one disabled refusal each

   37 caught      all 37 by the unit tests
   75 SURVIVED    undetected by the test suite AND the tamper sweep

MUTATION SCORE   37/112 = 0.330

survivors by refusal class
   26  raw-aggregate-mismatch       6  summary-mismatch
   20  artifact-unparsable          4  stamp-mismatch
   16  schema-violation             1  check-artifact-not-listed
                                    1  duplicate-artifact
                                    1  issuer-commit-mismatch
\end{Verbatim}

Two thirds of this verifier's refusals could be silently deleted with every gate still green.

The second number is worse than the first, and it needs stating more carefully than an
earlier draft of this paper managed. The 16 tamper fixtures exist as a CI job named
\texttt{invalidation-liveness}, whose stated purpose in the workflow file is that ``[t]he
verifier must prove it can both pass and BLOCK \dots{} A gate that never fires certifies
nothing.'' Of the 75 mutants the unit tests missed, that job caught \textbf{zero}.

That zero is real but it is not a surprise, and presenting it as one would be this paper
committing its own subject. It is entailed. At \texttt{f59fb62} every one of the 16 fixtures
is named in \texttt{tests/test\_verify.py}'s \texttt{TAMPERS} map, which asserts both
\texttt{verify\_bundle(FIX / name) == expected} and \texttt{main([...]) == 1} for each. So a
mutant that stops a fixture being refused fails a unit test first, the sweep returns
\texttt{tests}, and the fixture disjunct is never evaluated. Given a suite that pins every
fixture's exact verdict, the corpus's \emph{marginal} contribution over that suite is zero
by construction, not by misfortune.

The number that is not entailed is the corpus scored on its own. Re-running the same 112
mutants with the unit-suite disjunct removed, so that only the liveness control and the
fixture corpus can fire, gives \textbf{10 caught of 112, a score of 0.089} (measured
2026-08-18 on a clean copy of \texttt{f59fb62}). Ten of the sixteen fixtures each flip
exactly one mutant from refused to accepted; the other six flip none.

That 0.089 sits against a ceiling the corpus cannot exceed, and the ceiling is structural
rather than a matter of how well the fixtures are written. Each fixture is a single boolean
detector: it either exits 1 or it does not. If its bundle's failures all trace to one
refusal site, deleting that site flips it and deleting any other site does not; if they
trace to more than one, no single deletion flips it at all. So each fixture can be flipped by
at most one mutant in the whole population, and $n$ fixtures cannot score above $n/112$:
here $16/112 = 0.143$. Adding fixtures buys coverage linearly while refusal sites accumulate
with the code. A gate built to prove a verifier can refuse, sized at one fixture per imagined
attack, cannot keep up with a refusal population seven times its size. That is an argument
about the shape of hand-written corpora, not about this one's quality.

An honest caveat, and it is the one a reviewer should press hardest: disabling a refusal is
meaningful only if some input would have reached it, so some fraction of the 75 could be
unreachable or genuinely redundant rather than untested. Reachability is not established
per-survivor at this baseline.

What bounds that concern is not an analogy but this paper's own endpoint. By
\texttt{92e4548} a killing bundle had been constructed for every non-excluded site: the
sweep scores $143/143$, and a mutant dies only if some input reaches the deleted statement
and some assertion notices. Over the whole exercise exactly two sites turned out to be dead
code and three are excluded as unreachable (Sections~\ref{sec:exclusions}
and~\ref{sec:liveness-tests}), so the unreachable population is five enumerated lines rather
than an unknown fraction. The limit of that argument, stated so it is not overread: the 143
sites at \texttt{92e4548} are not the same set as the 112 at \texttt{f59fb62}, so this
bounds the concern rather than proving reachability survivor by survivor at the baseline.
Four independently confirmed forgeries separately establish that at least some baseline
survivors were live.

For external calibration in the other direction, static C analyzers miss 47\% to 80\% of
real vulnerabilities in benchmark programs \citep{lipp2022static}, and Beer et al.\ report
that roughly a fifth of passing formulas in industrial hardware verification were vacuous
\citep{beer2001vacuity}; a two-thirds gap looks like personal negligence in isolation and
like an industry baseline beside those.

\subsection{RQ4: does the operator anticipate the manual findings?}

This is the method's strongest validation, and it needs one correction the earlier draft did
not make.

\begin{table}[ht]
\centering
\caption{Each hand-found forgery against the surviving refusal population.}
\label{tab:predict}
\begin{tabular}{llll}
\toprule
Section & Forgery & Related survivor class & Anticipated? \\
\midrule
\ref{sec:sec-forge-summary} & string-typed summary lie & \texttt{summary-mismatch} (6) & yes \\
\ref{sec:sec-forge-check}   & deleting the check       & \texttt{check-artifact-not-listed} (1) & no, see below \\
\ref{sec:sec-forge-severity}& capital-letter severity  & \texttt{raw-aggregate-mismatch} (26) & yes \\
\ref{sec:sec-forge-stamp}   & deleting the stamp       & \texttt{stamp-mismatch} (4) & yes \\
\bottomrule
\end{tabular}
\end{table}

Three of the four hand-found forgeries make a \emph{live} refusal unreachable, and in each
case that refusal is a survivor: Section~\ref{sec:sec-forge-summary} makes a
\texttt{summary-mismatch} unreachable, Section~\ref{sec:sec-forge-severity} makes a
\texttt{raw-aggregate-mismatch} never fire, and Section~\ref{sec:sec-forge-stamp} makes a
\texttt{stamp-mismatch} skip its comparison. For those three the measurement does not merely
agree with the manual audit, it points at it.

The fourth is different and the earlier draft's table overstated it. The defect in
Section~\ref{sec:sec-forge-check} is an \emph{absent} obligation: no refusal site existed for
an evidence artifact covered by no check, and the fix had to create one. An operator that
deletes existing refusal statements cannot, by construction, see a refusal that was never
written. The surviving \texttt{check-artifact-not-listed} refusal is adjacent (it names the
inverse relation, a check referring to an artifact not listed) but it is not the same
obligation. Counting it as a prediction would be the paper indicting itself.

So the claim is: the operator anticipated three of four, and was blind to the fourth in a way
that is a property of the operator family rather than of this run. That is a weaker claim
than the draft made and it is still the strongest validation in the paper, because the four
forgeries were found independently and by hand.

The manual search took an external audit to teach me the shape, then a day of adversarial
work. The mutation run is unsupervised and cheap: at \texttt{92e4548} the 143-mutant sweep
completed in 5 minutes 46 seconds on an Apple M4 (measured once, 2026-08-18, on a clean copy of
that commit). The duration of the original 112-mutant run was not archived, and an
earlier draft's ``eleven minutes'' is not reproducible; it is dropped rather than restated.
What the run added beyond the four categories was 72 further places to look: of the 75
survivors, three are the refusals that three of the four hand-found forgeries make
unreachable, and the fourth subtracts nothing, because no refusal site existed for it.

\subsection{RQ2: what bug-specific hardening bought}
\label{sec:rq2}

I fixed all four defects in Section~\ref{sec:forgeries} and added a tamper fixture per fix,
so the \texttt{invalidation-liveness} job could prove each new refusal fires. Every forgery is
now refused by name, and all 20 fixtures are refused against a passing live control (verified
at \texttt{92e4548}: \texttt{fixtures/valid} exits 0 and all 20 \texttt{fixtures/tamper-*}
exit 1). Re-measuring at \texttt{6b6f96f}:

\begin{Verbatim}[frame=single,fontsize=\small]
BEFORE   37/112 = 0.330      (f59fb62, 114 tests, 16 fixtures)
AFTER    39/119 = 0.328      (6b6f96f, 114 tests, 20 fixtures)
\end{Verbatim}

The score did not move. Two of the four new fixtures kill a mutant; the other two kill none.
Closing four holes required seven new refusal statements, themselves untested. The
\texttt{stamp-mismatch} survivors went from 4 to 8. Four bugs fixed, seven guards added, net
coverage flat.

An earlier draft hedged that ``much of'' the decline was denominator growth. The archived
sweeps decompose exactly, and the exact statement is stronger than the hedge. The unit
suite caught 37 mutants before and 37 after; the two additional catches after are both by
new fixtures (\texttt{tamper-summary-string} and \texttt{tamper-check-deleted}). Caught
counts per refusal class are identical in every class except the two that gained a catch,
and in both of those the catch falls on a newly added site. So \textbf{the effect of the four
fixes on the 112 pre-existing sites is exactly zero}: that population scores $37/112$ both
before and after. The whole of the movement is the seven new sites, which were themselves
covered at $2/7$, a rate no better than the code being fixed.

The precise claim is therefore narrower than the headline number invites, and sharper:
\emph{the bug-specific regressions closed the four demonstrated vulnerabilities, left the
liveness of every pre-existing refusal untouched, and installed seven new refusals that
were themselves mostly untested}. They sampled the failures that had been discovered rather
than exercising the remaining population of refusals. Reported as a decimal delta the
change is $-0.003$ ($0.330357 \rightarrow 0.327731$); the rounded scores in the box above
differ by $-0.002$, and the fractions are quoted throughout in preference to either.

It would be wrong to say the fixes bought nothing. They closed four working forgeries and
pinned each against return. What they did not do, and what a careful engineer would expect
them to do, is make the \emph{rest} of the gates any more likely to fail when they should.
The mass was 26 surviving \texttt{raw-aggregate-mismatch} refusals concentrated in one
profile's checker, which no audit had a reason to look at.

\subsection{The measurement lied first, and scored a perfect 1.000}
\label{sec:vacuous1000}

The first post-hardening run reported \textbf{119/119 = 1.000}.

It was false. The harness ran \texttt{pytest -x}, and the baseline was \emph{already red}:
the two tests that assert the real evalmut and crashkit bundles verify clean had just been
(correctly) broken by the new evidence-unchecked rule. Pytest therefore exited nonzero for
every mutant, every mutant scored ``caught'', and the score reported a perfect suite while
measuring nothing at all.

A tool built to detect checks that pass without checking produced a check that passed without
checking, and the failure presented as the best possible result. Had I reported 1.000 it
would have been the most flattering and most worthless number in this project. The published
literature has the same warning from the other side: assertion-free suites still achieve
substantial mutation scores by relying on uncontrolled implicit checks
\citep{schuler2011checkedcov}.

The fix is the rule this paper already argues for, applied to itself, and it is the validity
gate of Section~\ref{sec:gate}. A liveness gate on the liveness instrument.

\subsection{RQ3: what did move it, testing the refusals rather than the bugs}
\label{sec:liveness-tests}

The survivors were never a mystery. They were a work list. Seventy-five tests were written,
one per refusal surviving at the 112-site baseline, organised by cluster rather than by
defect. See Section~\ref{sec:host} for a provenance caveat on how those tests were produced,
which matters more than the count.

The discipline that made them worth anything is the same one the rest of this paper is about.
Each test asserts the \emph{exact} failure list, never a substring search, so it cannot pass
on a cascade it did not cause. And for most of them, disabling the target refusal makes
\texttt{verify\_bundle} return \texttt{[]} and the cooked bundle verifies clean, which proves
that refusal is the \emph{sole} guard for its defect rather than one voice in a chorus.
Several tests were rewritten mid-flight when that check revealed they were leaning on a
stale-hash backstop instead of the arithmetic they claimed to pin. This is oracle improvement
in the sense of \citet{jahangirova2016oracle}, and the mechanism by which it works is the one
\citet{zhang2015assertions} measured: what moves detection is assertions on behaviour, not
more executed code.

Table~\ref{tab:ladder} gives the full ladder with the code state each row was measured on,
because the two headline deltas are otherwise not comparable: the 0.941 run sits on the
post-fix tree, so it already contains the seven refusal sites the fixes added.

\begin{table}[ht]
\centering
\caption{Refusal-site liveness coverage by code state. Sites and test counts measured
directly at each commit; scores read from the archived sweep outputs in \texttt{paper/}.
The 0.330 and 0.328 rows use no exclusions. The fourth row's commit was not recorded
alongside its archived sweep at the time and is recovered from the repository: eight
commits carry 134 refusal sites, but only three of them carry a one-line exclusion set and
so can produce the denominator 133, and of those \texttt{a17af5c} is ruled out because it
scored $123/133$ (Section~\ref{sec:host}). The remaining two differ only in paper files, and
\texttt{1674f4c} is the commit that introduced \texttt{paper/mutation\_final2.json}, 42
seconds after that file's own mtime. The row is pinned there.}
\label{tab:ladder}
\begin{tabular}{llrrrl}
\toprule
Stage & Commit & Sites (excl.) & Caught & Score & Tests \\
\midrule
baseline                          & \texttt{f59fb62} & 112 (0) & 37  & 0.330 & 114 \\
after the four bug fixes + fixtures & \texttt{6b6f96f} & 119 (0) & 39  & 0.328 & 114 \\
after per-refusal liveness tests  & \texttt{6c3a2c8} & 119 (0) & 112 & 0.941 & 189 \\
after closing every issuer's gap  & \texttt{1674f4c} & 134 (1) & 126 & 0.947 & 217 \\
after covering the remainder      & \texttt{92e4548} & 146 (3) & 143 & 1.000 & 258 \\
\bottomrule
\end{tabular}
\end{table}

\paragraph{The full record, including the parts the table smooths.} Table~\ref{tab:ladder}
shows five rows: one fall and then three clean rises. The repository holds twelve archived
sweeps, two of which record the same $39/119$, so eleven values, and they fall three more
times. In commit order the measured scores are 0.330, 0.328, 0.941, 0.957, 0.960,
0.953, 0.947, 0.992, 1.000, 0.972 and 1.000. The score fell four times: $0.330 \rightarrow
0.328$ when the four bug fixes added seven sites, $0.960 \rightarrow
0.953 \rightarrow 0.947$, and $1.000 \rightarrow 0.972$ when a new profile added twelve
refusal sites. In every one of those falls the numerator rose or held (the sequence of
caught counts, 37, 39, 112, 112, 120, 123, 126, 130, 131, 139, 143, never decreases); what
moved was the denominator. Nothing ever stopped being caught. New gates arrived untested.

That is the RQ2 phenomenon recurring \emph{after} the RQ3 intervention, and it is the
strongest argument in this paper for making the sweep a standing CI floor rather than a
one-time measurement: a codebase that adds refusals adds them uncovered by default, and the
score decays unless something asks. Reporting only the table's five rows would have made a
sawtooth look like a ramp, which is a presentation choice this paper in particular cannot
afford.

Two further clarifications the table forces, and both were errors in the earlier draft.
First, the 75 tests were one per survivor at the \emph{112-site} baseline; by the time they were measured
the tree had 119 sites and 80 survivors, so ``one per surviving refusal'' describes the work
list, not the tree the score was taken on. Second, the $0.330 \rightarrow 0.941$ and
$0.330 \rightarrow 0.328$ deltas are both quoted against the same baseline for comparability,
but the second measurement is taken on top of the first, not as an alternative to it.

At the 0.947 row, seven refusals still survived, and the honest breakdown mattered: two were
dead code, not untested. One was unreachable because a non-dict \texttt{flips.json} is already
refused upstream by the JSON loader's \texttt{want=dict}; the other was an \texttt{OSError}
wrapper no bundle-shaped input can reach, since the artifact must already have been read and
hashed to enter the check. Both were probed empirically rather than assumed, and no test was
written to fake coverage of them. The dead one was deleted; the unreachable one moved into the
exclusion set described in Section~\ref{sec:exclusions}, which has since grown to three lines
across two fragments as more render comparators were added.

The 1.000 at \texttt{92e4548} is not the vacuous 1.000 of Section~\ref{sec:vacuous1000}: the
sweep's own liveness gate passed first, printing \texttt{baseline clean; 143 refusal sites (3
excluded, see EXCLUDE)}. I reproduced it independently on a clean copy of that commit, and it
matches the project's own continuous-integration log for the same commit. CI enforces
\texttt{-{}-floor 0.99}.

\paragraph{How much of the $+0.611$ is entailed.} Most of it, and saying so is the price of
the rest of the paper. The RQ3 intervention is defined in the metric's own units: one test
per surviving site, each asserting the exact failure list that site produces. A test shaped
that way fails when its site is deleted, by construction. So once the tests land, the score
moves; the movement measures completion of a work list, not an independent gain in oracle
quality. The RQ2 arm, by contrast, was defined over \emph{defects} and never targeted a
site, which is what any defect-shaped intervention scores on a site-liveness metric. The
gap between $-0.002$ and $+0.611$ therefore measures targeting at least as much as efficacy,
and no design in this paper separates the two.

Two things in that run are not entailed, and they are the section's real content. The first
is reachability: it was not known in advance that a bundle-shaped input existed for each
site, and two sites turned out to have none (Section~\ref{sec:liveness-tests}). The second
is that the resulting metric has teeth on code written afterwards, which is a claim about
the future rather than about the work list, and Section~\ref{sec:host} is where it was
tested: the CI floor caught a comparator I had added and not tested, at a commit where no
test had been written for it. Earning the efficacy claim properly needs the pre-registered
second operator family (Section~\ref{sec:fixes-open}), scored against tests that were not
written with that operator in view. That is not done.

With those qualifications, the comparison between the two middle rows is still the paper's
practical claim. Fixing named defects and pinning each with a regression fixture, which is
the instinctive post-audit response, left the pre-existing population at $37/112$.
Systematically asking ``can this gate fail?'' of every gate took it to $112/119$.
\textbf{The defects you find are a sample; the gates you own are the population.} Offered as
a heuristic drawn from one system, not as a demonstrated law, and partly true by
construction.

\subsection{The closure rule found the same gap in three issuers}
\label{sec:closure}

The rule from Section~\ref{sec:sec-forge-check}, that an artifact listed as evidence must be
read by some check, was written to close one forgery. Turned on the live registry, it refused
three of the five issuer families.

\begin{table}[ht]
\centering
\caption{Artifacts pinned as evidence but read by no check, when the closure rule was first
run against the live registry.}
\label{tab:closure}
\begin{tabular}{ll}
\toprule
Issuer & Pinned but unchecked \\
\midrule
crashkit      & \texttt{variance\_flaky\_n10.report.json} \\
evalmut       & \texttt{dogfood\_gradecore.txt}, \texttt{promptfoo\_findings.txt} \\
agent-certlab & \texttt{CONTRACT.md}, in all seven certifications \\
\bottomrule
\end{tabular}
\end{table}

Three of the four artifacts are the human-readable view: two \texttt{.txt} renders and a
\texttt{.md} contract. The fourth does not fit that description and is recorded rather than
smoothed over: crashkit's \texttt{variance\_flaky\_n10.report.json} is machine-readable JSON
and was unbound anyway. So the class the rule actually caught is broader than
``human-readable'': it is \emph{artifacts produced as reports rather than consumed as a
check's input}. Under that description all four fit, and the human-readable ones are the
subset where it costs most, because that is the document a person actually reads. Crashkit's
capability sentence claimed its variance report ``aggregates reproducibly''; certlab's
contracts announced ``N/M seeded defects fixed''. True claims, unverified.

The fix in each case was to bind the render to its payload. Not byte-identity re-rendering,
which would couple the verifier to another repository's formatting, but holding the render's
headline to the recomputed values, with an unparseable render refused rather than skipped.

I do not claim the other two issuer families were clean. The rule fired on three; whether
fleet-board and modeldrift had no closure gap, or simply were not reached by this sweep, is
not established here.

\paragraph{Stated as a hypothesis, because the sample cannot support more.} These are repos
by one author, in one suite, under one set of conventions: a within-author architectural
pattern, not evidence about attacker behaviour or about the industry. What it suggests, and
what is testable, is that \emph{an artifact emitted as a report is systematically underbound
relative to an artifact consumed as a check's input}. The mechanism is mundane enough to be
plausible elsewhere: authors bind what a check reads and treat what a reader reads as
disposable, right up until the second is the thing people act on. The crashkit row is the
case that constrains the wording, since it is a report in JSON: what predicts the gap is the
artifact's role, not its file format, and a narrower ``renders go unbound'' hypothesis is
already refuted at $n=4$. Zhao et al.'s
population-scale catalogue of leaderboard smells is the nearest independent evidence that the
pattern is not unique to me \citep{zhao2025lbops}, though I have not checked their smell
definitions against this mechanism and do not claim the mapping.

Testing it properly means going outside this codebase: sample public attestations,
CI-generated security reports, evaluation write-ups and model cards; ask whether the displayed
summary is deterministically derived from signed or bound data; separate generated reports
from hand-authored narrative; and fix the coding rule before looking at outcomes. Fourn\'e et
al.'s interview study is the methodological template for doing that properly
\citep{fourne2023flossing}. That study is not in this paper.

\paragraph{Present state, stated honestly.} At the 2026-08-16 measurement all three issuers
were re-emitted, landed clean, and the registry re-pinned byte-identically. That is no longer
true. Measured on 2026-08-18 at \texttt{92e4548}: crashkit verifies with 0 failures; evalmut
fails with 1, an \texttt{evidence-unchecked} on \texttt{dogfood\_fixtures.json} and
\texttt{promptfoo\_fixtures.json}; model-drift fails with 14, thirteen
\texttt{raw-aggregate-mismatch} rows in \texttt{standings.json} plus a render mismatch. The
verifier has not changed since \texttt{ba14203} (2026-08-16); the issuer bundles drifted. Notably,
evalmut's re-emission reintroduced exactly the gap this rule exists to catch, which is a live
datapoint in the rule's favour and an uncomfortable one for the project's release hygiene.

\subsection{Host dependence of the measurement}
\label{sec:host}

The CI floor failed on its first real push. At \texttt{a17af5c}, the \texttt{refusal-coverage}
job printed \texttt{baseline clean; 133 refusal sites (1 excluded)}, then \texttt{MUTATION
SCORE: 123/133 = 0.925} and \texttt{FAIL: 0.925 is below the floor 0.940}. Part of that was
my own omission: I had added a comparator and not tested it, which is exactly the rot the
floor exists to catch, and it caught me.

An earlier draft compared that 0.925 against ``0.953 locally'' and attributed the gap to host
dependence. That comparison was wrong and I am reporting the correction rather than quietly
dropping it. The 0.953 is $123/129$, measured at \texttt{bfc9642}, a tree with 130 refusal
sites. The CI figure is $123/133$ at \texttt{a17af5c}, a tree with 134. The intervening commit
added exactly the four refusal sites that account for the entire difference, and the numerator
is identical at 123. Nothing stopped being caught. The gap was commit skew.

The underlying host-dependence claim is nevertheless true, and I re-established it by a
different route. Fourteen tests are gated on sibling issuer checkouts being present. Without
them the suite runs 238 passed, 16 skipped, 4 xfailed; with the checkouts correctly wired the
same suite runs 14 more tests. So the suite a developer runs and the suite CI runs are not the
same suite, which is the same class as a verdict that changes with the filesystem encoding,
the defect the external audit opened with. What I can no longer support is the specific claim that
four mutants stop being caught. At \texttt{92e4548} the sweep scores 143/143 on a checkout
\emph{without} the sibling repositories, which is the only configuration in which it currently
runs at all: with them present, two drifted issuer bundles make the baseline red and the sweep
aborts by design (Section~\ref{sec:gate}). So that claim is not re-measurable today, and it is
withdrawn rather than restated.

An earlier draft also stated that CI now checks out the issuers ``so it measures what a
developer measures''. It does not. That is instrument failure seven in
Section~\ref{sec:instruments}.

\paragraph{Provenance of the 75 tests.} The 75 tests were not written serially by one
author, and the 0.941 above is a single serialised sweep rather than an aggregate of their
self-reports. The reason that distinction matters is a threat to validity and is stated under
Test provenance in Section~\ref{sec:threats}.

\section{Why the existing suite missed all of this}
\label{sec:whymissed}

Every fixture encodes a forgery I had already imagined. \texttt{tamper-wrong-sha256} tests a
hash that does not match its artifact; the audit's attack re-pinned the hash \emph{honestly}.
The fixture set is a map of my own threat model, and a threat model cannot contain its own
blind spot by construction. The obvious external correction is to derive fixtures from a
published attack taxonomy \citep{ladisa2023sok} or to generate them adversarially rather
than by hand, as frankencerts do for certificate validation
\citep{brubaker2014frankencerts}.

The 114 tests were not idle either. They executed the verifier extensively. What they did not
do is assert that its refusals could fire, which is the distinction checked coverage was
invented to measure \citep{schuler2011checkedcov} and the reason coverage is a poor quality
target \citep{inozemtseva2014coverage}. A covered line is not an asserted line, and 75
surviving refusal sites is what that gap looks like when you count it.

The sharper point concerns a tool I had already built and never applied here. \emph{evalmut}
injects known defects into a system and reports which checks stayed green. \vacname{}
verifies evalmut bundles; it had never been mutated by evalmut. The verifier is a grader, and I never graded the grader. That is Thompson's regress
in its most mundane possible form \citep{thompson1984trusting}, and it is where the auditor
walked in.

\section{Instrument failures observed during the study}
\label{sec:instruments}

This section exists because omitting it would make the paper an instance of its own subject.

\begin{enumerate}
\item A tamper sweep printed \emph{all 16 refused} while every invocation had exited 127. A
startup failure, not a refusal. The loop scored ``nonzero exit'' as ``correctly refused''.
\item Re-running the sweep months later, all 16 exited 2. A wrong module path. Identical
symptom, different cause. The live control caught it; without the control the run would have
read as a clean sweep.
\item Inside the script I wrote to hunt this bug, a probe edit failed on a wrong filename and
printed \texttt{exit=0}. That was the \emph{unmodified} bundle passing. The tool built to find
vacuous passes produced one.
\item A proof-of-concept for a fifth finding ran against a stale local \texttt{origin/main}.
After \texttt{git fetch} the premise evaporated. The finding was void.
\item The first post-hardening mutation run scored a perfect 1.000 against an already-red
baseline: \texttt{pytest -x} exited nonzero for every mutant, so all 119 scored ``caught''.
The tool built to find vacuous passes produced one, and it presented as the best possible
result (Section~\ref{sec:vacuous1000}).
\item Editing \emph{this paper} to insert the mutation result, a \texttt{str.replace()} on the
abstract matched nothing and returned the string unchanged. The build succeeded, the PDF
regenerated, and the abstract still carried the old numbers. Python's \texttt{str.replace}
cannot fail; it can only decline to do anything. The fix was to switch to an editor that
errors on no-match, which is the entire thesis of this paper applied to a text edit.
\item The remediation written to remove host dependence reports success along a path where the
thing it added is never read. The workflow sets \texttt{VAC\_EVALMUT\_CHECKOUT} and its two
siblings to \texttt{\_issuers/<repo>/vac}, but the three consumers
(\texttt{tests/test\_verify.py:286}, \texttt{:325} and \texttt{:349} at \texttt{92e4548})
append \texttt{/vac} themselves, so the value resolves to
\texttt{\_issuers/<repo>/vac/vac}. That directory does not exist in any of the three issuer
repositories, whose bundle manifest sits at \texttt{<repo>/vac/vac.json}. Each
\texttt{skipif} therefore fires and the six tests gated on those three variables skip inside
the very job that checked the repositories out. The other eight issuer-gated tests skip for an
unrelated reason: \texttt{VAC\_CERTLAB\_CHECKOUT} and \texttt{VAC\_FLEET\_CHECKOUT} are never
set by the workflow at all, so agent-certlab's seven registry entries and reference-fleet's one
were never in reach. Two independent causes, one silent zero, and the totals hide the seam.
Nothing in the log gives it away, which is the point: the
\texttt{refusal-coverage} job emits no pytest tally at all, because the sweep captures
pytest's output and prints only its own lines. What does give it away is that the job's score
at \texttt{92e4548} is $143/143$, identical to a sweep run with the variables unset. An
earlier draft cited the \texttt{238 passed, 16 skipped, 4 xfailed} line as the evidence; that
line is emitted by the \texttt{test (3.11)} and \texttt{test (3.12)} jobs, which never check
the issuers out and were never meant to, so it could not evidence anything about
\texttt{refusal-coverage}. Citing the wrong job's output for a claim about instrument blindness is a reporting error
rather than an instrument failure: no tool returned success without measuring, I cited the
wrong measurement. It is recorded here rather than silently repaired, and the underlying
defect stands on the path arithmetic above, not on the line I misquoted.
\end{enumerate}

Seven instrument failures in the system under study, all producing plausible output, plus one
reporting error in this paper's own account of the seventh. Four of the seven happened
inside instruments written specifically to detect this class, across three distinct tools: the
tamper sweep (items 1 and 2), the probe script (item 3), and the mutation sweep itself
(item 5). Two of them (items 1 and 2) are CWE-390 exactly: an error condition detected and
converted into a success signal \citep{cwe390}. The most dangerous was not the one that broke.
It was the one that returned a perfect score.

Items 3 and 6 also share a root cause with two of the errors this paper corrects. The wrong
refusal-site idiom and the wrong test count in the earlier draft both came from a working copy
of a commit that had been reverted, which is the same failure as item 4: measuring a tree that
was not the tree under test.

The operational rules that survive:

\begin{itemize}
\item \textbf{Prove the instrument before the finding.} A liveness control adjacent to every
sweep.
\item \textbf{Re-run the ruled-out list after any fix.} A stale exclusion is
indistinguishable from a real one.
\item \textbf{Prefer operations that fail loudly over operations that silently no-op.}
\texttt{str.replace}, \texttt{dict.get(k, default)}, \texttt{if k in a and k in b}, and
\texttt{grep \textbar{} true} are the same hazard wearing different clothes: they convert ``did not
happen'' into ``fine''.
\end{itemize}

\section{Discussion}
\label{sec:discussion}

Any eval suite, CI gate, or verifier can be audited with two questions.

\begin{enumerate}
\item \textbf{Can this check pass without executing?} Empty input, a missing file, a process
that fails to start, a pattern that matches nothing, a swallowed exception, an unknown enum
value absorbed by a default. This is the CWE-703 family \citep{cwe703}, and its fail-open
variant has a name and a number \citep{cwe636}.
\item \textbf{Has the detector been proven able to fire, in this run, on this host?} Not ``it
has a test''. A liveness control adjacent to the assertion.
\end{enumerate}

A gate that has never been observed failing has not been observed working. That is the whole
argument for mutation-testing eval suites rather than trusting their green, and it is the
argument the sanity-checking literature made first for model checkers
\citep{kupferman2006sanity}.

The one caution worth adding is against a grep. At \texttt{f59fb62},
\texttt{grep -rn severity tests/} returned zero, which was true and useful. At
\texttt{92e4548} it returns 13, because the fix brought tests. A pattern that matches nothing
is evidence only if you have separately established that it \emph{can} match, which is the
same hazard the third operational rule above names.

\section{Scope of claims}
\label{sec:scope}

Six things, and nothing broader.

\begin{enumerate}
\item A reproducible case study of five fail-open and vacuous-pass classes in an offline
evidence-bundle verifier, each with a working forgery: four found against the hardened
verifier (Section~\ref{sec:forgeries}) and one, the \texttt{null}-artifact forgery, reported
by the external audit against the pre-hardening version (Section~\ref{sec:audit}).
\item A refusal-site liveness instrument: the deletion operator, the enumerated refusal-site
denominator, the deterministic two-detector kill predicate, the exclusion set, and the
validity gate. The operator family, the practice of grading a checker by mutation, and the
green-baseline precondition are all prior art (Sections~\ref{sec:related-neighbours}
and~\ref{sec:related-unclaimed}); the site selection and the subject are what is offered,
and the kill predicate is what makes the measurement decomposable rather than what makes it
new.
\item An empirical before-and-after on \emph{one} real verifier, comparing discovered-bug
regression testing against systematic refusal-site liveness testing.
\item A separately scored detector result: the committed tamper corpus catches 10 of 112
refusal sites on its own, against a structural ceiling of $16/112$
(Section~\ref{sec:rq1}).
\item A finding-shaped \emph{hypothesis} that an evidence artifact emitted as a report is
systematically underbound relative to one consumed as a check's input
(Section~\ref{sec:closure}), untested outside this codebase.
\item A record of seven instrument failures during the study
(Section~\ref{sec:instruments}).
\end{enumerate}

It does not claim a general result about post-audit engineering practice, about verification
systems broadly, or that this methodology transfers. It does not claim that pointing a
mutation tool at a checker is new, that reading surviving mutants as blind spots in a checker
is new, or that a continuously green suite shipping defects is a new observation.
Section~\ref{sec:related-neighbours} names who reported each.

\textbf{Every number here is self-measured on a system I wrote.} The verifier, the tests, the
mutation operator, the exclusion set, and this paper are all mine. The registry it verifies is
a closed loop: all entries are from my own repositories across five of my own projects, which
is one multi-repository fixture rather than five independent issuers. The single external data
point in the entire record is the audit in Section~\ref{sec:audit}. This is stated prominently
rather than in a footnote because a paper about instruments that report success without
checking has no business burying the fact that it graded itself.

\section{Threats to validity}
\label{sec:threats}

\subsection{Construct validity}
\label{sec:limits-score}

\paragraph{What the score is and is not.} The operator disables refusal statements only. It
does not mutate comparison operators, boundaries, or control flow. So the number is not an
upper bound on anything: an upper bound constrains a broader unknown, and this constrains
nothing outside its own population. It is a \emph{complete score against a deliberately narrow
operator over an enumerated set of refusal sites}, in the adequacy sense of
\citet{zhu1997adequacy}, and 1.000 means that obligation is discharged, not that the verifier
is correct. \citet{just2014mutants} bound the substitution from the other side: 17\% of real
faults are coupled to no mutant generated by common operators at all.

\paragraph{A live counterexample to what 1.000 means.} The strongest evidence that the score
does not measure correctness arrived while this paper was being written, from outside it. Pull
request 8 reports a covered artifact replaced by a symlink pointing outside the bundle: the run
exits 0 and prints \texttt{structural verification: PASS}, because the bytes hash identically,
they simply live somewhere else on the verifying host. A verifier scoring 1.000 on this paper's
operator accepts that bundle. The two facts are compatible rather than contradictory, and their
compatibility is the point: the operator asks whether each refusal that exists can fire, and a
refusal nobody wrote has no site to delete. Section~\ref{sec:fixes-open} carries it as open.
The honest reading of 1.000 is that the enumerated obligation is discharged, and nothing more.

\paragraph{The denominator is a source-level proxy.} A refusal site is one \texttt{f.append(}
statement. That misses rejections expressed as raises, early returns, assertions, or exit-code
propagation, and reasons built indirectly through helpers. The omission is countable rather
than abstract: at \texttt{92e4548} six rejection points sit outside the mutated population and
are never mutated, three early returns (\texttt{verify.py:1508} missing-manifest,
\texttt{:1512} and \texttt{:1514} invalid-json) and three \texttt{raise ValueError}
unsafe-tar-member rejections (\texttt{:1541}, \texttt{:1546}, \texttt{:1558}). The same six
are outside it at \texttt{f59fb62}. More importantly the proxy misses fail-open behaviour that
happens \emph{before} any refusal is reached: parser defaults, canonicalisation, decoding,
path resolution, duplicate-key handling, exception swallowing. And, per
Section~\ref{sec:gate}, the population is enumerated rather than pinned, so it can move under
a refactor without the gate objecting. The honest name for what is measured is
\textbf{refusal-append liveness coverage}, and it is the name that should be read wherever
this paper says ``mutation score'', including at 1.000.

\paragraph{The mutants themselves are sound.} The obvious suspicion about a deletion operator
is that the score is inflated by mutants that fail to compile or that kill the process rather
than an assertion. Checked at \texttt{92e4548}: all 146 mutants parse (\texttt{ast.parse}, 0
failures), no statement span swallows a neighbouring refusal site, and 115 of the 146 statements
are multi-line, which the paren-balance span handles. A random sample of eight mutants
(seed 7) all died on assertion failures comparing \texttt{verify\_bundle} output against an
expected failure list, none on an unhandled exception. That sample was run on CPython 3.14
rather than the project's 3.11 and 3.12, with one pre-existing byte-identity failure
deselected, so it is a check on the operator rather than a figure from the paper's own
environment.

\paragraph{Closure is artifact-read coverage, not field-binding coverage.} The rule in
Section~\ref{sec:sec-forge-check} proves a check \emph{references} an artifact; nothing in it
proves the check reads anything inside. A check could open a file and bind none of it.
\texttt{tests/test\_refs\_are\_bound\_not\_just\_read.py} probes this mechanically by
corrupting each referenced artifact and re-pinning honestly. Measured at \texttt{92e4548}:
13 refs, of which 2 carry no corruptible value and are skipped, leaving 11 actually corrupted.
Of those 11, four still verified clean. Two were the \texttt{schema} format-version field, one
a non-load-bearing digit in a raw stream, and one an unrecomputed \texttt{models} count inside
a narrative. These are unbound \emph{fields} inside bound artifacts, not decorative refs, and
they are recorded in a \texttt{KNOWN\_UNBOUND} inventory. An earlier draft said ``4 of 12'',
which was wrong in both the denominator and the composition; the same error is in that test
file's own docstring, which is worth noting because the paper was quoting the code and the code
was quoting nothing. The three-level ladder this exposes, artifact-read to field-binding to
claim coverage, is roadmap, not result.

\subsection{Internal validity}

\paragraph{The author designed the operator and holds the result.} I wrote the verifier, chose
the mutation operator, defined the kill predicate, wrote the exclusion set, and wrote this
paper. The operator could in principle have been selected after seeing which sites survived.
The control I propose for this is pre-registration of a second operator family
(Section~\ref{sec:fixes-open}) before running it, which is the practice the mutation-testing
methodology literature recommends \citep{papadakis2019mutation}. It is not yet done, so this
threat is open.

\paragraph{Test provenance.} The 75 liveness tests were produced by six concurrent workers
sharing a single working tree, and at least one of them observed a sibling mutating
\texttt{vac/verify.py} underneath its own verification sweep. Their self-reported
``mutation-checked'' counts are therefore not independently trustworthy: the tree each worker
measured was not the tree it wrote against. The claim is the measurement, not the authorship.
The 0.941 is a single serialised run of \texttt{tools/mutation\_sweep.py} against a clean tree
with the liveness gate satisfied.

\paragraph{The instruments failed seven times.} Section~\ref{sec:instruments} is, read
strictly, an internal-validity section. Four of the seven failures occurred inside the
measurement apparatus, and one produced a perfect score. Every number in this paper should be
read as carrying that risk, mitigated but not eliminated by the liveness gate.

\paragraph{Reachability is not established per survivor.} The 0.330 counts surviving refusal
sites, not surviving \emph{reachable} refusal sites. Two dead refusals were found and handled
explicitly (Section~\ref{sec:liveness-tests}); the rest were not individually probed at the
baseline.

\subsection{External validity}

One verifier, one author, 11 registry entries across five repositories all owned by me. There
is no second issuer, no second codebase, and no second annotator. The one external data point
is the audit. The closest thing to independent corroboration of the \emph{class} is
CVE-2022-35929 in cosign \citep{cve2022sigstore}, which is a different system by different
authors exhibiting the same shape.

\subsection{Conclusion validity}

There is no control condition, and the two arms differ in more than one way at once. The move
from $37/112$ to $112/119$ is attributable to per-refusal liveness testing, but it is not
separable from the effort expended: 75 tests is a large intervention, and a comparison against
75 tests spent some other way was not run. The null result for bug-specific hardening is on a
much smaller intervention (four fixes, four fixtures), so the two arms are not effort-matched.
They are also not matched in \emph{targeting}, which is the larger problem and is stated in
full in Section~\ref{sec:liveness-tests}: the RQ3 intervention is defined over the metric's own
units and the RQ2 one is not, so part of the gap is definitional rather than empirical.
Additionally, the score is host-dependent in the sense of Section~\ref{sec:host}: which tests
run depends on which sibling repositories are on the machine.

\section{Fixes}
\label{sec:fixes}

\subsection{Landed}

\begin{itemize}
\item \textbf{Unknown severity is a named refusal}, not weight 0. The lookup indexes strictly
and a label outside the frozen table is \texttt{artifact-unparsable}, named by the offending
label. \texttt{SPEC.md} changed with it and records the replaced rule explicitly.
\item \textbf{An evidence artifact not covered by any check is a named refusal},
\texttt{evidence-unchecked}.
\item \textbf{Non-numeric summary values are compared, not skipped.} A JSON string that parses
as a number is now \texttt{summary-outruns-checks}. \texttt{SPEC.md} was amended rather than
patched around, and it records why.
\item \textbf{Stamp keys named in \texttt{protocol.hashes} but absent from the artifact are a
\texttt{stamp-mismatch}}, matching the two profiles that already failed closed.
\item \textbf{A tamper fixture per fix.} 20 fixtures, all refused against a passing live
control.
\item \textbf{The mutation sweep is a CI floor.} \texttt{tools/mutation\_sweep.py -{}-floor
0.99} runs in the \texttt{refusal-coverage} job.
\item \textbf{The two unreachable refusals are resolved.} One was deleted; one moved into the
documented exclusion set. This bullet appeared as a recommendation in the earlier draft, citing
\texttt{verify.py:863} and \texttt{:990} at \texttt{27809ce}; it is done.
\end{itemize}

\subsection{Outstanding}
\label{sec:fixes-open}

\begin{itemize}
\item \textbf{Bind \texttt{acc} to its own evidence.} Reported by the auditor on 18 August
2026 and reproduced here. The specification never ties a series point's \texttt{acc} to its
\texttt{fails} vector, and \texttt{verify.py:1242} enforces the relation only for the
\texttt{mock:*} control series. For every real series the declared accuracy is bound to no
evidence at all. In \texttt{fixtures/valid} the series \texttt{beta:b1} declares 0.66 and 0.9
where its own \texttt{fails} vectors imply materially lower values, while \texttt{alpha:a1} and
\texttt{gamma:g1} happen to be consistent. This is the closure rule of
Section~\ref{sec:closure} one level down: there an artifact was listed but unread, here a field
is read but unbound. It sits inside the bundle this paper calls valid, and the verifier scoring
1.000 on the operator of Section~\ref{sec:method} accepts it.

\item \textbf{Close the symlink escape, reported after this study closed.} Pull request 8,
open at the time of writing, reports that a covered artifact replaced by a symlink pointing
outside the bundle exits 0 and prints \texttt{structural verification: PASS}. The bytes hash
identically because they are the same bytes; they live somewhere else on the verifying host.
That is a fifth vacuous pass against the hardened verifier, of the same class as
Section~\ref{sec:forgeries} and outside the refusal-site population every number in this paper
is measured over. It is deliberately not landed here: it changes \texttt{vac/verify.py}, and
therefore the site count, the score, and every figure pinned to \texttt{92e4548}. Landing it
belongs to the next measurement, not to this one, and saying so is cheaper than a paper whose
numbers quietly stopped matching its own repository.

\item \textbf{Fix the CI issuer-checkout paths} so the env values do not compose to a doubled
\texttt{vac/vac} directory, and add a hard assertion that no test skips with the reason
``issuer checkout not present'', so a path that silently resolves to nothing fails loudly
instead of measuring a smaller suite. See Section~\ref{sec:instruments}, item 7.
\item \textbf{Re-emit the drifted issuer bundles} so the registry and the suite are green
again (Section~\ref{sec:closure}).
\item \textbf{Get one external issuer.} Five same-author repositories are one fixture, not
five issuers. A single outsider who builds a bundle from their own workflow, hits a real
refusal or a specification ambiguity, and either lands or publishes the blocker is worth more
than any further internal profile.
\item \textbf{Climb the coverage ladder}: artifact-read (done) to field-binding (probed, gaps
recorded) to claim coverage, where every externally visible claim names its evidence source and
the verifier's obligation over it.
\item \textbf{Pre-register a second operator family} before running it: comparison inversion,
boundary shifts, missing-key guard removal, enum and default perturbation, so the result is not
chosen after seeing the score.
\item \textbf{Cross-platform determinism in CI} (Linux, macOS, Windows), at minimum locale,
encoding and path semantics. The encoding bug in Section~\ref{sec:audit} was found by a
stranger, not by me.
\item \textbf{Derive tamper fixtures from a published attack taxonomy}
\citep{ladisa2023sok} rather than from imagination, since a threat model cannot contain its
own blind spot.
\item \textbf{Land the robustness pull request.} It is open and has been reworked in response
to a revert. The reverted version built two fixtures with
\texttt{json.loads('\{"a":' * 3000)}, which raises \texttt{RecursionError} inside the
\emph{decoder} before the verifier is called, so the tests were green on one interpreter and
red on another: the same host-dependence class the patch exists to remove. At
\texttt{ce462e10}, the head of the pull request on 2026-08-18 and resolvable only through
\texttt{git fetch origin pull/2/head} because the branch sits on a fork, those tests build
their structure in Python instead, and both issuer-data traversals are iterative with explicit
stacks, so the stated blocker is met by the pull request itself. The head carries a date
because it moved during drafting: the earlier \texttt{bb4dda36}, which this bullet named
first, still asserted a wall-clock bound on elapsed time, an instance of the same
host-dependence class; \texttt{ce462e10} drops that assertion and leaves \texttt{vac/}
byte-identical. What remains is review and merge. An earlier draft of this paper still
described the blocker as open, having been written against a stale reading of the branch;
that is instrument failure 4 recurring inside the paper's own outstanding-work list, which is
why it is corrected here in place rather than dropped.

\item \textbf{Pin the refusal-site population} the way \texttt{EXCLUDE} pins its arity, so a
refactor that merges refusal sites aborts the sweep instead of shrinking the denominator
(Section~\ref{sec:gate}). Written rather than outlined: \texttt{tools/mutation\_sweep.py} now
declares \texttt{EXPECT\_RAW\_SITES} and \texttt{EXPECT\_SCORED\_SITES} and aborts with exit 2
when the measured population differs, with \texttt{-{}-expect-sites} for a deliberate
measurement at another revision. The gate was checked by declaring a population one smaller
than the source carries and confirming the run aborts, rather than by assuming a written
constant is an enforced one. Landed at \texttt{e932868}.

\item \textbf{Commit the standalone-detector mode} used for the $10/112$ figure in
Section~\ref{sec:rq1}. \emph{Done.} \texttt{tools/mutation\_sweep.py} takes
\texttt{-{}-detector} and \texttt{tools/fixture\_corpus\_score.py} scores the corpus at
\texttt{f59fb62}, where the sweep does not exist; both landed at \texttt{e932868}, and
\texttt{fixtures/attack-crashkit-severity} (Table~\ref{tab:fixtures}) at \texttt{b30ea2f}.
Every disjunct of the kill predicate can now be scored on its own from a clone. This bullet is
kept rather than deleted because an earlier draft of it promised work that did not exist, and
the correction is more useful to a reader than a tidy list.
\end{itemize}

\section{Conclusion}

I asked four questions of one verifier and got four answers. Two thirds of its refusal sites
could be deleted with every gate still green (RQ1: 75 of 112, score 0.330). Regression testing
the four defects an external audit and a day of manual search had produced left the caught set
on those 112 sites at exactly 37, and installed seven new refusals covered at $2/7$ (RQ2).
Writing one liveness test per surviving refusal took the score to $112/119$, and covering the
remainder reached 1.000 against this operator, though the size of that second move is partly
entailed by defining the intervention in the metric's own units (RQ3). The operator's survivor
classification anticipated three of the four hand-found forgeries and was blind to the fourth
for a reason that is a property of the operator family, not of the run (RQ4).

The number I would most like a reader to take is not 1.000. It is 0.089. That is what the
sixteen-fixture corpus behind the continuous-integration job named
\texttt{invalidation-liveness}, whose stated purpose is to prove the verifier can refuse,
scores as a detector standing on its own: 10 of 112. Measured the way the sweep actually runs
it, downstream of a unit suite that already asserts every fixture's exact verdict, its
marginal contribution is a flat 0, and that 0 is arithmetic rather than misfortune. Either
number says the same thing. A corpus sized at one fixture per imagined attack cannot cover a
refusal population seven times its size, and a gate that has never been observed
failing has not been observed working. I built the gate to observe failure and never asked it
to observe most of them.

The single thing that would most change the picture is not another operator or another metric.
It is one external issuer: somebody who is not me, building a bundle from their own workflow,
hitting a refusal I did not anticipate.

\section*{Acknowledgements}

Giulio D'Erme ran the audit that started this, in four public pull requests against the
artifact repository. The four findings his first review did not reach were only findable
because that review taught me the shape to look for; the encoding defect he found was one I
had no route to discovering alone; and the symlink case recorded in
Section~\ref{sec:audit} is his as well, found after the hardening, in the class this paper
is about. Any errors that remain are mine, including,
on the evidence of Section~\ref{sec:instruments}, some I have not yet detected.

\section*{Reproducibility and artifacts}

The verifier, the specification, the mutation sweep, the fixture corpus, the registry, and the
archived sweep outputs are public under the MIT licence at
\url{https://github.com/egnaro9/vac-protocol}.

This paper is itself archived and citable, independently of any preprint server:
\textbf{DOI \href{https://doi.org/10.5281/zenodo.22018308}{10.5281/zenodo.22018308}}, a concept
DOI resolving to the newest version, with \texttt{10.5281/zenodo.22018309} pinning this one. It
is a preprint deposit and carries no peer review, no moderation, and no endorsement; it
establishes that this text existed on 2026-08-19 and nothing more.

The state of that repository at the time of writing is archived and citable:
\textbf{DOI \href{https://doi.org/10.5281/zenodo.22000911}{10.5281/zenodo.22000911}}, which
resolves to the most recent archived release. Release \texttt{v0.1.0}, deposited 2026-08-18 and
built from commit \texttt{11291be}, carries its own version DOI
\href{https://doi.org/10.5281/zenodo.22000912}{10.5281/zenodo.22000912}. Cite the version DOI
to pin a number in this paper to the exact code that produced it; the concept DOI is for the
project rather than for a result. Every quantitative claim here names a commit for the same
reason: a result belongs to the code that produced it, not to whatever \texttt{main} holds
later.

\paragraph{Commits.} Check out \texttt{92e4548} for every present-day figure in this paper.
The pre-fix baseline for Section~\ref{sec:forgeries} and the 0.330 measurement is
\texttt{f59fb62}. The intermediate ladder rows in Table~\ref{tab:ladder} name their own
commits. Two line citations in the earlier draft's fix list resolve only at \texttt{27809ce},
which is noted where they appear.

\paragraph{Environment.} Python 3.11 and 3.12 (the versions the project's CI matrix uses).
Install with \texttt{pip install -e ".[test]"}. Clone with \texttt{core.autocrlf} disabled: at
\texttt{92e4548} the repository carries no \texttt{.gitattributes}, and a checkout that rewrites
line endings makes the commands below fail, or worse, refuse for the wrong reason. The last
paragraph of this section records what that costs, because it is not a footnote.

\paragraph{Commands.}
\begin{Verbatim}[frame=single,fontsize=\small]
python -m pytest tests/ -q                       # 258 collected at 92e4548
python -m vac.verify fixtures/valid              # liveness control, exits 0
for d in fixtures/tamper-*; do \
    python -m vac.verify "$d"; done              # each exits 1
python tools/mutation_sweep.py --floor 0.99 --json out.json
grep -cE '^[[:space:]]*(f|failures)\.append\(' vac/verify.py   # 146
\end{Verbatim}

\paragraph{Reproducing the four forgeries.} Each is pinned by a committed fixture, and all
four are now one command in both directions. The third reaches its pre-fix direction through a
second fixture, which the caption names along with the one condition still attached to it.

\begin{table}[ht]
\centering
\caption{Forgery to fixture mapping. Each fixture exits 1 at \texttt{92e4548} with the named
reason. Copied into an \texttt{f59fb62} checkout, the first, second and fourth exit 0, which
is the pre-fix forgery reproduced in one command. The third takes a second fixture:
\texttt{tamper-crashkit-severity} re-cases three severities but leaves
\texttt{vulnerability\_score} declared at the honest $0.4545$, so the pre-fix verifier refuses
it on the ordinary \texttt{raw-aggregate-mismatch} path, and it pins the post-fix refusal only.
\texttt{fixtures/attack-crashkit-severity} completes the forgery: the same three re-casings,
with $0.0$ declared in \texttt{metrics.vulnerability\_score}, in
\texttt{results.summary.crash\_vulnerability} and in the \texttt{crashkit-battery-v1} check's
\texttt{expect} block, and \texttt{evidence/eval\_run.json} re-pinned honestly. It exits 0 at
\texttt{f59fb62} and exits 1 at \texttt{92e4548} on \texttt{artifact-unparsable}, both
re-measured on 2026-08-18 against a liveness control that passes at each revision. That fifth
fixture landed at \texttt{b30ea2f}, two commits after \texttt{92e4548}, so it is absent from
both revisions named here and has to be copied into a checkout of either; from
\texttt{b30ea2f} onward, including the archived \texttt{v0.1.0}, it is present in a clean
checkout, with \texttt{vac/} unchanged across that span (Section~\ref{sec:fixes-open}).}
\label{tab:fixtures}
\begin{tabular}{lll}
\toprule
Section & Fixture & Refusal reason at \texttt{92e4548} \\
\midrule
\ref{sec:sec-forge-summary}  & \texttt{fixtures/tamper-summary-string}     & \texttt{summary-outruns-checks} \\
\ref{sec:sec-forge-check}    & \texttt{fixtures/tamper-check-deleted}      & \texttt{evidence-unchecked} \\
\ref{sec:sec-forge-severity} & \texttt{fixtures/tamper-crashkit-severity}  & \texttt{artifact-unparsable} \\
\ref{sec:sec-forge-stamp}    & \texttt{fixtures/tamper-stamp-deleted}      & \texttt{stamp-mismatch} \\
\bottomrule
\end{tabular}
\end{table}

\paragraph{A warning that is itself a finding.} Reproduction is host-dependent. Fourteen tests
are gated on sibling issuer checkouts, and \texttt{vac/registry.py} names five variables, not
the three the workflow sets: \texttt{VAC\_CRASHKIT\_CHECKOUT}, \texttt{VAC\_EVALMUT\_CHECKOUT},
\texttt{VAC\_MODELDRIFT\_CHECKOUT}, \texttt{VAC\_CERTLAB\_CHECKOUT} and
\texttt{VAC\_FLEET\_CHECKOUT}, each pointing at a repository \emph{root}, since the consumers
append \texttt{/vac} themselves. With none of them bound the suite runs 238 passed, 16 skipped,
4 xfailed; with all five it runs 14 more. Binding only the first three runs 6 more, not 14, and
nothing in the output says so, which is item 7 of Section~\ref{sec:instruments} in a second
costume: a checkout variable that resolves to nothing measures a smaller suite quietly. At the
time of writing two of the issuer bundles fail, and both are issuer drift.
\texttt{model-drift} fails identically at \texttt{f59fb62} and at \texttt{92e4548} on the same
bundle bytes, so no change in the verifier accounts for it. \texttt{evalmut} passes at
\texttt{f59fb62} and fails at \texttt{92e4548} on \texttt{evidence-unchecked}, a refusal that
does not exist in the \texttt{f59fb62} source, which reads like verifier tightening until the
baseline is chosen correctly. \texttt{f59fb62} predates that rule entirely. The commit that
matters is \texttt{ba14203}, the 2026-08-16 state at which evalmut last landed clean: the rule
is already present there, and \texttt{git log ba14203..92e4548 vac/verify.py} is empty, so
the verifier has not moved since. Measured against that baseline evalmut is bundle drift too,
which is what Section~\ref{sec:closure} reports. Either way the baseline is red and the mutation sweep aborts by
design. The 143/143 figure is therefore reproducible on a checkout \emph{without} the sibling
repositories, which is how CI measures it and how I re-measured it on 2026-08-18.

\paragraph{A default Windows clone breaks this reproduction, and lies about how.} The commands
above assume a checkout that preserves bytes. At \texttt{92e4548} the repository carries no
\texttt{.gitattributes} (\texttt{git ls-tree 92e4548 .gitattributes} is empty),
while \texttt{core.autocrlf} is enabled by default by the Git for Windows installer, so a clone
made there rewrites every text artifact to CRLF at checkout and the manifests go on hashing raw
bytes. Measured on a differential pair of clones at \texttt{92e4548} differing only in
\texttt{core.autocrlf}: all fourteen files under \texttt{fixtures/valid} carry CR in the enabled arm
and none do in the disabled arm, and the liveness control exits 1 with thirteen
\texttt{sha256-mismatch} reasons rather than exiting 0. Thirteen is the ceiling for this bundle,
since \texttt{fixtures/valid} lists thirteen evidence entries and \texttt{vac.json} is not its
own evidence.

The second half is this paper's own thesis, occurring inside its reproduction section. All
twenty \texttt{tamper-*} fixtures still exit 1 on that clone, so the loop above still passes and
reports nothing wrong. But fourteen of the twenty no longer produce the refusal they were
written to demonstrate: the hash mismatch fires and the original reason is gone. Three of
Table~\ref{tab:fixtures}'s four rows name a reason the verifier no longer gives, with every exit
code that a reader would check still matching. Only \texttt{tamper-check-deleted} still emits
\texttt{evidence-unchecked}, and it does so buried among fourteen reasons. A loop over exit codes
is a check on the exit code and not on the refusal, and it stayed green through exactly the
substitution this paper is about, in the section that tells a stranger how to confirm the paper.

The checkout defect was filed as pull request 9 and merged at \texttt{81f50cf}, which adds a
one-line \texttt{.gitattributes} (\texttt{* -text}); re-measuring the same differential pair at
that commit gives exit 0 in both arms. That fix does not reach the artifacts this paper pins. It
is absent at \texttt{92e4548}, the commit every figure here is measured at, and absent from the
\texttt{v0.1.0} archive (\texttt{git ls-tree v0.1.0 .gitattributes} is empty). Until those are
re-pinned, disabling \texttt{core.autocrlf} at clone time is the only thing between a reader and
a verifier that refuses honest bundles and misnames dishonest ones. I did not find the checkout
defect; it was found by someone cloning the repository the way a stranger would, which is the
only way it was ever going to be found from a project whose nine CI jobs all run
\texttt{ubuntu-latest}. The second half, that the tamper loop keeps passing while the reasons
change underneath it, surfaced only on re-running that loop on such a clone and reading the
reasons instead of the exit codes.

\paragraph{Archived measurements.} The sweep outputs backing Table~\ref{tab:ladder} are
committed under \texttt{paper/} as JSON: \texttt{mutation.json} ($37/112$),
\texttt{mutation\_after.json} ($39/119$), \texttt{mutation\_covered.json} ($112/119$),
\texttt{mutation\_final2.json} ($126/133$), and \texttt{mutation\_v8.json} ($143/143$). Seven
further sweeps are archived in the same directory and are the source of the non-monotone
record in Section~\ref{sec:liveness-tests}: \texttt{mutation\_gated.json} ($39/119$),
\texttt{mutation\_final.json} ($112/117$), \texttt{mutation\_v2.json} ($120/125$),
\texttt{mutation\_v3.json} ($123/129$), \texttt{mutation\_v5.json} ($130/131$),
\texttt{mutation\_v6.json} ($131/131$) and \texttt{mutation\_v7.json} ($139/143$). Each file's
introducing commit is recoverable with \texttt{git log -{}-diff-filter=A -{}- paper/<file>},
which is how the fourth row of Table~\ref{tab:ladder} was pinned.

\paragraph{The standalone fixture-corpus score.} The $10/112$ in Section~\ref{sec:rq1} is not
one of the archived sweeps, and it is not produced by \texttt{tools/mutation\_sweep.py}. It is
produced by a second script, committed at \texttt{e932868}, in one command:

\begin{Verbatim}[frame=single,fontsize=\small]
python tools/fixture_corpus_score.py                    # 10/112 = 0.089
python tools/fixture_corpus_score.py --detector liveness # 0/112 = 0.000
\end{Verbatim}

The script creates and removes its own detached worktree at \texttt{f59fb62}, refuses to run
against a checkout sitting at any other revision or carrying uncommitted changes, and applies
the same operator over the same \texttt{span()} logic, imported from
\texttt{tools/mutation\_sweep.py} rather than copied. Only the sixteen \texttt{tamper-*}
directories may report a catch, and the ten catches are credited to ten distinct fixtures. The
second line scores the clean-bundle control the same way and catches nothing, so none of the
ten is borrowed from it. The second script is necessary rather than untidy:
\texttt{mutation\_sweep.py} now takes \texttt{-{}-detector fixtures}, which is the right way to
ask this of the tree as it stands, but that file does not exist at \texttt{f59fb62}, and
pointing HEAD's copy at a checkout of it aborts by design, since its \texttt{EXCLUDE} map is
keyed to HEAD's source and the arity guard refuses to resize the denominator rather than
quietly measure a different one. Both the script and the \texttt{-{}-detector} flag landed at
\texttt{e932868} (Section~\ref{sec:fixes-open}), so this figure is re-runnable in one command
from a clone rather than only on the author's tree. Both numbers above were re-measured on
2026-08-18 from a fresh clone of the public repository.

\bibliographystyle{plainnat}
\bibliography{refs}

\end{document}